\documentclass[prl,twocolumn]{revtex4-2}

\usepackage{}
\usepackage{epsfig}
\usepackage{graphicx}
\usepackage{amsfonts,amssymb}
\usepackage{amsmath}
\usepackage{color}
\usepackage {times}
\usepackage{mathrsfs}
\usepackage{hyperref}
\usepackage{comment}

\definecolor{refcolor}{rgb}{1.0,0.0,0.0}

\begin{document}

\title{Disorder enhanced ferromagnetic polaron formation \\
- and the test case of Europium Oxide}

\author{Tanmoy Mondal$^1$ and Pinaki Majumdar$^2$}

\affiliation{$^1$~Harish-Chandra Research Institute 
(A CI of Homi Bhabha National Institute), 
Chhatnag Road, Jhusi, Allahabad, India 211019\\
$^2$~School of Arts and Sciences, Ahmedabad University, 
Navrangpura, Ahmedabad,
India 380009}
\date{\today}

\begin{abstract}
Europium Oxide (EuO), a low carrier density local moment ferromagnet, shows a 
wide variety of transport behaviour depending on preparative conditions. Some 
samples have a moderate resistivity with a modest peak near $T_c$ while others 
show a huge peak in resistivity followed by insulating high temperature 
behaviour. These features have been known for decades and have been attributed 
to the presence of magnetic polarons in a disordered background. Actual attempts
at a theory, however, reduce the problem either to a single trapped electron 
or to an averaged picture where the spatial physics of polarons is lost. The
difficulty stems from having to handle electronic states in a magnetically
fluctuating, structurally disordered background. Via an explicit real space
calculation in two dimensions, we examine the interplay of disorder induced 
localisation and magnetic polaron formation and show how the resistivity trends 
in EuO could emerge from increasing impurity concentration. We estimate the 
polaron size in the disordered medium, establish the presence of a pseudogap 
near $T_c$, predict a crossover to incoherent, non Drude, optical response with 
growing disorder and temperature, and track the polaron `delocalisation' with 
increasing magnetic field.  
\end{abstract}

\maketitle

\section{Introduction}

Electrons in non disordered low carrier density ferromagnets are 
delocalized in the low temperature ordered state but can ``self-trap'' 
as the magnetic disorder increases with increasing temperature.
This self-trapping arises from a competition between internal 
energy gain and entropy loss in localising the electrons in a 
small neighbourhood, locally polarising the spins. The self 
trapped object is called a ferromagnetic polaron (FP)
\cite{ref_EuS,ref_other_Eu_materials,ref_exp2,ref_exp4,ref_EuB6_opt_cond1}. 
The most common indication of polaron physics is in the temperature 
dependent resistivity $\rho(T)$. In polaronic materials, the magnetic 
scattering induced rise of $\rho(T)$ with increasing temperature, and 
saturation above $T_c$, is replaced by a peak near the ferromagnetic 
$T_c$, followed by a wide window with $d\rho/dT < 0$ . 
This non monotonic resistivity has been observed in a variety of 
low carrier density local moment ferromagnets 
\cite{ref_EuB6_resistivity,ref_other1,ref_other_Eu_materials,
ref_exp2,ref_exp3, ref_exp_GdSi,ref_exp4,ref_EuB6_opt_cond1}.

Low carrier density, and the possibility of self-trapping, make 
structural disorder an important player in these materials. 
Defects generated in the synthesis process, or introduced through 
doping, can serve as nucleation centers for electron trapping
\cite{ref_EuO_res_1,ref_EuO_res_2,ref_GdN_res,
ref_GdN_dope,ref_GdN_band1,ref_GdN_band2,ref_GdN_gap}. 
A prominent example in this class is EuO, a local-moment ferromagnet 
with $S = 7/2$ moments on Eu, carrier densities $n_{el} \sim 10^{-3} - 
10^{-2}$ per formula unit, and and $T_c \sim 70$K. EuO exhibits a 
strongly sample dependent resistivity, which is primarily attributed 
to variation in carrier density and impurity concentration 
\cite{ref_EuO_res_1,ref_EuO_res_2,ref_EuO_map, 
ref_EuO_split,ref_EuO_band1,ref_EuO_band2}. 
This arises because of the chemistry of EuO,
it can easily oxidise to Eu$_2$O$_3$ under 
oxygen-rich conditions, while oxygen-poor environments favour the 
formation of oxygen vacancies.  Even slight deviations from precise 
oxygen pressure can lead to defect formation \cite{ref_EuO_dope}.

This leads to dramatic variation in the resistivity in samples 
that are nominally labeled EuO. Even if we ignore the strongly
insulating samples, which probably have very low carrier density 
\cite{ref_EuO_res_1, ref_EuO_res_2},
the relatively `metallic' samples display the following behaviour: 
(i)~the $T \rightarrow 0$ resistivity spans a couple of decades, 
from $10^{-5}~\Omega$cm to $10^{-3}~\Omega$cm,  
(ii)~the resistivity at $T_c$ varies from $10^{-3}~\Omega$cm to 
$10~\Omega$cm, with a $d \rho/dT < 0$ window above $T_c$ but no 
hint of activated behaviour, (iii)~the magnetoresistance can be 
huge, $\sim 50\%$ at $100$ mTesla and $\sim 90\%$ at $1$ Tesla. 
To create a contrast, in EuB$_6$, another much studied material,
one observes a nearly sample independent resistivity, with 
$\rho(0) \sim 10^{-6}~\Omega$cm 
and $\rho(T_c) \sim 3 \times 10^{-4}~\Omega$cm, 
a very shallow minimum in the $T > T_c$ resistivity, 
and only modest magnetoresistance \cite{ref_EuB6_res_field,ref_EuB6_hall, ref_EuB6_stm}. 

Although there are polaronic effects in EuB$_6$ as well,
it is believed that polaronic trends are amplified in EuO 
due to defect related trapping centers. Indeed, evidence from 
Raman scattering \cite{ref_EuO_raman} and muon spin relaxation 
\cite{ref_EuO_meuon} suggests that defects extend the polaronic
temperature window in EuO compared to the cleaner EuB$_6$. 
However, there is only limited understanding of the spatial 
physics of magnetic polarons, in a many electron 
situation, in the presence of impurities.

Previous theoretical studies of EuO have tried to address the 
resistivity by incorporating impurity effects at various levels 
of approximation
\cite{ref_EuO_theory_res_1,ref_EuO_theory_res_2,
ref_EuO_theory_res_3,ref_EuO_theory_res_4,
ref_EuO_theory_res_5,ref_EuO_theory_res_6,
ref_EuO_theory_res_7}.
All of these start with a model of ferromagnetically coupled Eu 
spins, with conduction electrons coupled to the Eu moments via 
$s–f$ exchange and moving in the background of dilute impurities.
One approach is to use a variational scheme
\cite{ref_EuO_theory_res_1,ref_EuO_theory_res_2,ref_EuO_theory_res_3},
to compute the wavefunction of a single  electron trapped around 
an impurity, and compute the `binding energy' as a function of 
impurity potential, temperature and magnetic field. Typically, 
the temperature and magnetic field $h$ enter through their effect 
on the magnetisation $m(T,h)$. This is the `bound magnetic polaron' 
(BMP) scenario \cite{ref_EuO_theory_res_2}, and the binding energy 
that emerges from here is used as an activation gap in the 
conductivity. Impurities are just traps, there is no treatment of 
`disorder' within this scheme, or thermal fluctuations. A more 
sophisticated approach 
considers a multi-impurity situation but then treats the problem 
through a coherent potential approximation (CPA)  
\cite{ref_EuO_theory_res_4,ref_EuO_theory_res_5,
ref_EuO_theory_res_6,ref_EuO_theory_res_7}
where again the conductivity is controlled by the
spectral gap that emerges.
The approach has had success in reproducing some of the
observed features of EuO but (i)~it predicts an activated
$T > T_c$ conductivity inconsistent with experiments, and
(ii)~it reveals little about polarons and their spatial 
physics.

In this paper, we set a twofold task. First, we want to clarify 
the general interplay of structural disorder and magnetic
thermal fluctuations in trapping electrons in polaronic
states, and contrast the spatial, transport, spectral
and optical features with that in the `clean' problem.
Second, we use parameter values aligned closely with that
in EuO, and vary only one parameter, the impurity density, 
to see how much of EuO physics can be captured, and what 
features, not yet measured, can be predicted.

We adopt the generally accepted model
\cite{ref_theo_other_montecarlo,ref_theory_1polaron,
ref_theo_other_variation,ref_theo_other_dmft,
ref_theory_magneto_resistance,
ref_theory_polaron_hopping,ref_EuB6}
involving a lattice of Heisenberg spins with ferromagnetic
coupling $J$, with the spins ``Kondo'' coupled to itinerant
electrons with coupling $J'$. The electrons have a tight
binding hopping $t$ and experience an impurity potential 
$V$ at a fraction of sites. We work in two 
dimensions (2D) but make some remarks on the 3D case.
The key difficulty in this problem is in computing
the effect of the electrons on the core spins, at finite
temperature. We use a Langevin dynamics (LD) scheme 
 \cite{ref_EuB6,ref_spin_lang,ref_spin_lang1,ref_chern} 
to treat thermal fluctuation of the spins, with an exact 
diagonalisation method to compute the torque. 
An alternate Monte Carlo approach to these models has 
typically been limited to size $\sim 10 \times 10$ but the
parallelised LD approach can access  lattices with 
$N \sim 20 \times 20$. This size, though still modest, allows 
us to study a finite concentration of polarons and map their 
transport, spatial, and spectral features.

To keep the study manageable, we set parameters based on
prior experimental 
\cite{ref_EuO_split,ref_EuO_band1,ref_EuO_band2} 
and theoretical \cite{ref_EuO_theory_res_4,
ref_EuO_theory_res_5,ref_EuO_theory_res_6,ref_EuO_theory_res_7}
studies on EuO,
setting, $J'S = t$, and $JS^2 = 0.01t$ and $V=-2t$, and 
electron density $n_{el} = 2.5 \%$. We only vary the impurity 
concentration $n_{imp}$ as our measure of disorder and present 
our results for varying $\eta = n_{imp}/n_{el}$.
Our main results are the following.

(i)~Resistivity:
We observe increasing nonmonotonicity in the  resistivity 
with increasing $\eta$. While the overall resistivity increases
with disorder, the two ratios, $\alpha_1= \rho(T_c)/\rho(0.5T_c)$ 
and $\alpha_2 = \rho(T_c)/\rho(2T_c)$, increase rapidly.
Our 2D approach overestimates $\rho(0.5T_c)$, so underestimates
$\alpha_1$, but $\alpha_2$ is in excellent agreement with data.
Also, while the high temperature phase is `insulating' 
it does not have an activated form, consistent
with the lower resistivity EuO samples 
\cite{ref_EuO_res_1,ref_EuO_res_2}.

(ii)~Spatial textures:
While disorder leads to an inhomogeneous electron density
at all $\eta$ and $T$, polaronic textures are observed only
over a window $\sim 0.8T_c-1.5T_c$,
depending on $\eta$.  The area $A_p$ associated with a 
polaron, inferred from the spatial maps, decreases from  
$A_p \sim 25$ sites in the clean limit to $A_p \sim 10$ sites 
at $\eta=2$. The linear spread of the polaron is $\sim 
\sqrt{A_p}$.

(iii)~Density of states: The $T \ll T_c$ density of states is 
featureless for the $\eta$ values we use, but a pseudogap (PG) 
appears at the chemical potential at a disorder dependent temperature.
This PG formation temperature reduces, and the 
depth of the PG increases, with increasing $\eta$. 
Also, an increase in $\eta$ pushes the PG closing 
temperature to a higher value. For our parameter values 
there is no clean gap at any $\eta$ and $T$.

(iv)~Optical conductivity: $\sigma(\omega)$ has a Drude character at 
small $\eta$ and low temperature but both increasing disorder and/or 
increasing temperature lead to a non Drude response, with the peak 
in $\sigma(\omega)$ shifting to a finite frequency. By $T \gtrsim 
2T_c$ the optical spectrum is completely incoherent.

(v)~Magnetoresistance: Using an applied magnetic field $h$ as 
another control parameter we find that in moderately disordered 
samples the magnetoresistance $(\rho(T,0) - \rho(T,h))/\rho(T,0)$
at $T \sim T_c$ is $\sim 50\%$ at $h/T_c \sim 0.06$  and $\sim 80 
\%$ at $h/T_c \sim 0.2$. These values are consistent with measurements 
on EuO, and much larger than what one observes in `clean' systems 
like EuB$_6$. The spatial maps indicate that by $h \sim 0.2 T_c$ 
the polarons have delocalised, while a pseudogap that was present at 
$h=0$ has disappeared, and the optics approaches Drude behaviour.

The rest of the paper is organised as follows. 
We describe the model and computational method in
the next section.  We then examine
the resistivity $\rho(T)$ that emerges for varying disorder,
followed by a look at spatial maps of the electron density and
ferromagnetic correlation for the same parameters. Following this,
we look at the global and local density of states, and the
optical conductivity to identify polaronic signatures.
Finally, we look at the `homogenisation' that happens
on applying a magnetic field, and the attendant
changes in transport and spectral properties. We
conclude by suggesting an interpretation of the
resistivity and it's possible dependence on
dimensionality.

\section{Model and Method}

We consider a square lattice with a spin 
of unit magnitude, at each site, exchange coupled to the nearest 
neighbours by a coupling $J$. The spins are additionally coupled 
locally to electrons via a coupling $J'$, and the electrons hop 
between nearest neighbour sites with an amplitude $t$.
\begin{eqnarray}
H &=& 
~~~~~ H_{kin}~~~~ + ~~~~H_{el-sp}~~~~ + ~~~~H_{dis}~~~~ + ~~~~H_{sp}\cr
\cr
&=&
 -t \sum_{ij}^{\sigma} c^{\dagger}_{i\sigma} c_{j\sigma}
- J'\sum_i {\bf S}_i.{\vec \sigma}_i 
+ \sum_i V_i n_i
-J \sum_{ij} {\bf S}_i.{\bf S}_j 
\nonumber
\end{eqnarray}
The moment of Eu is large, $7/2$, so we consider the spins 
$S$ as classical rotors of unit magnitude.
We set $t=1$ and represent other parameters in terms 
of it $JS^2/t = 0.01$, and $J'S/t =1.0$. 
We set the potential $V_i=V = -2t$ on a fraction $n_{imp}$ 
of randomly chosen lattice sites and zero elsewhere.
The electron density $n_{el}$ is set to $0.025$. 
Since the effective strength of disorder is set by
the ratio of the Born scattering rate $\Gamma = 
n_{imp}V^2 D(\epsilon_F)$
and $\epsilon_F$, where $\epsilon_F$ is the Fermi energy and
$ D(\epsilon_F)$ is the density of states at the Fermi level,
our effective disorder parameter 
is $\Gamma/\epsilon_F 
\propto (n_{imp}/n_{el}) V^2/t^2$ in 2D. Since $V/t$ is
fixed we define $\eta = n_{imp}/n_{el}$ as our effective 
measure of disorder.

There are two steps in solving the problem: (i)~generating the
spin configurations $\{{\bf S}_i \}$ appropriate to a set of
electronic parameters, impurity configuration, and temperature,
and (ii)~using the electronic eigenstates in the $\{{\bf S}_i \}$
background to compute transport and spectral properties. 
Part (i) requires evaluation of the spin distribution function
$$
P\{{\bf S}_i\} \propto
Tr_{el} \big[e^{-\beta(H_{kin} + H_{el-sp} + H_{dis} + H_{sp})} \big]
$$
The associated effective Hamiltonian for the spins is
\begin{eqnarray}
H_{eff}\{{\bf S}\} &=& - J \sum_{ij} {\bf S}_i.{\bf S}_j 
- T log \big[Tr e^{- \beta H_{el}} \big]
\cr
H_{el} &=& H_{kin} + H_{el-sp} + H_{dis}
\nonumber
\end{eqnarray}
where the `trace' is carried out over electronic states.
As the form above shows we only have an implicit form for $H_{eff}$
since the electronic states in an arbitrary spin and impurity 
background are unknown and can only be computed numerically at
strong electron-spin coupling. Equilibrium spin configurations
can be computed via a Metropolis algorithm that iteratively 
diagonalises $H_{el}$. 
Although the spins are classical, this iterative calculation
involves a $O(N^4)$ cost at a given temperature and limits system
size to $\sim 10 \times 10$. An equivalent dynamical scheme
(see below),
which computes the torque on the spins and uses thermal noise
to attain equilibriation, is more amenable to parallelisation 
and costs $O(N^3)$. We use this method and can access sizes
$\sim 20 \times 20$, adequate to study a multipolaron situation.

Within our Langevin dynamics (LD) approach, the spin
${\bf S}_i$ is
subject to a systematic torque ${\bf T}_i$ arising from the 
interaction with other spins and the electron moment, and a 
`random' torque ${\bf h}_i$ proportional to $\sqrt{T}$.
The Langevin equation 
\cite{ref_spin_lang, ref_spin_lang1, ref_chern} 
governing the evolution of spins is given by:
\begin{eqnarray}
\frac{d\mathbf{S}_i}{dt} &=& \mathbf{S}_i \times (\mathbf{T}_i
+ \mathbf{h}_i) - \gamma \mathbf{S}_i \times (\mathbf{S}_i
\times \mathbf{T}_i) \cr
\mathbf{T}_i &=& -\frac{\partial H}{\partial \mathbf{S}_i} =
-J' \langle \vec{\sigma}_i \rangle -
J \sum_{\langle j \rangle} \mathbf{S}_j
\cr
\langle h_{i \alpha} \rangle &=& 0,~~
\langle
 h_{i \alpha}(t)h_{j \beta}(t') \rangle = 2\gamma k_B T
 \delta_{ij} \delta_{\alpha \beta} \delta(t-t')
\nonumber
\end{eqnarray}
$\gamma$ is a damping constant (which does not affect
equilibrium properties), and is set to $0.1t$,  and 
$\mathbf{h}_i$
satisfies the fluctuation-dissipation theorem.
Here, $\langle \vec{\sigma}_i \rangle$ is 
the expectation of $\vec{\sigma}_i$ in the
instantaneous ground state of the electrons in
the spin configuration in the previous time step.

From the equilibrium spin configurations $\{{\bf S}_i^{\alpha}\}$,
where $\alpha$ is the configuration index, we compute the
electronic eigenstates $\psi_n^{\alpha}$ and eigenvalues
$\epsilon_n^{\alpha}$. From these, we construct the thermally
averaged electron density $n_i$, short range ferromagnetic 
correlation $C_i$, the density of states $D(\omega)$, the
optical conductivity $\sigma(\omega)$, and the d.c resistivity
$\rho(T,h)$. The expressions for these are given in the 
Appendix. We disorder average our $D(\omega)$ and 
$\sigma(\omega)$ over 
at least five impurity configurations.

\section{Resistivity}

Fig.1 shows our results for the temperature-dependent
 resistivity $\rho(T)$, alongside experimental data for EuO
 \cite{ref_EuO_res_1,ref_EuO_res_2}. In panel (a) the
 experimental $\rho(T)$ data for five different EuO samples 
\cite{ref_EuO_res_1,ref_EuO_res_2}
 all exhibit a pronounced peak at $T_c$, followed by a sharp 
 decrease.  Our calculated results of $\rho(T)$ are shown in 
panel (b) where we vary $\eta$.  Even for a low impurity
 density, $\eta = 0.5$, our resistivity exhibits an enhanced
 peak-dip feature compared to the clean, $\eta=0$, case.
As $\eta$ increases, the peak resistivity rises along with a
sharper fall after the peak. The increase in $\eta$ also
causes a change in the low temperature behaviour, making it
 insulating gradually \cite{footnote}.

\begin{figure}[b]
\centerline{
\includegraphics[height=7.3cm,width=8.5cm]{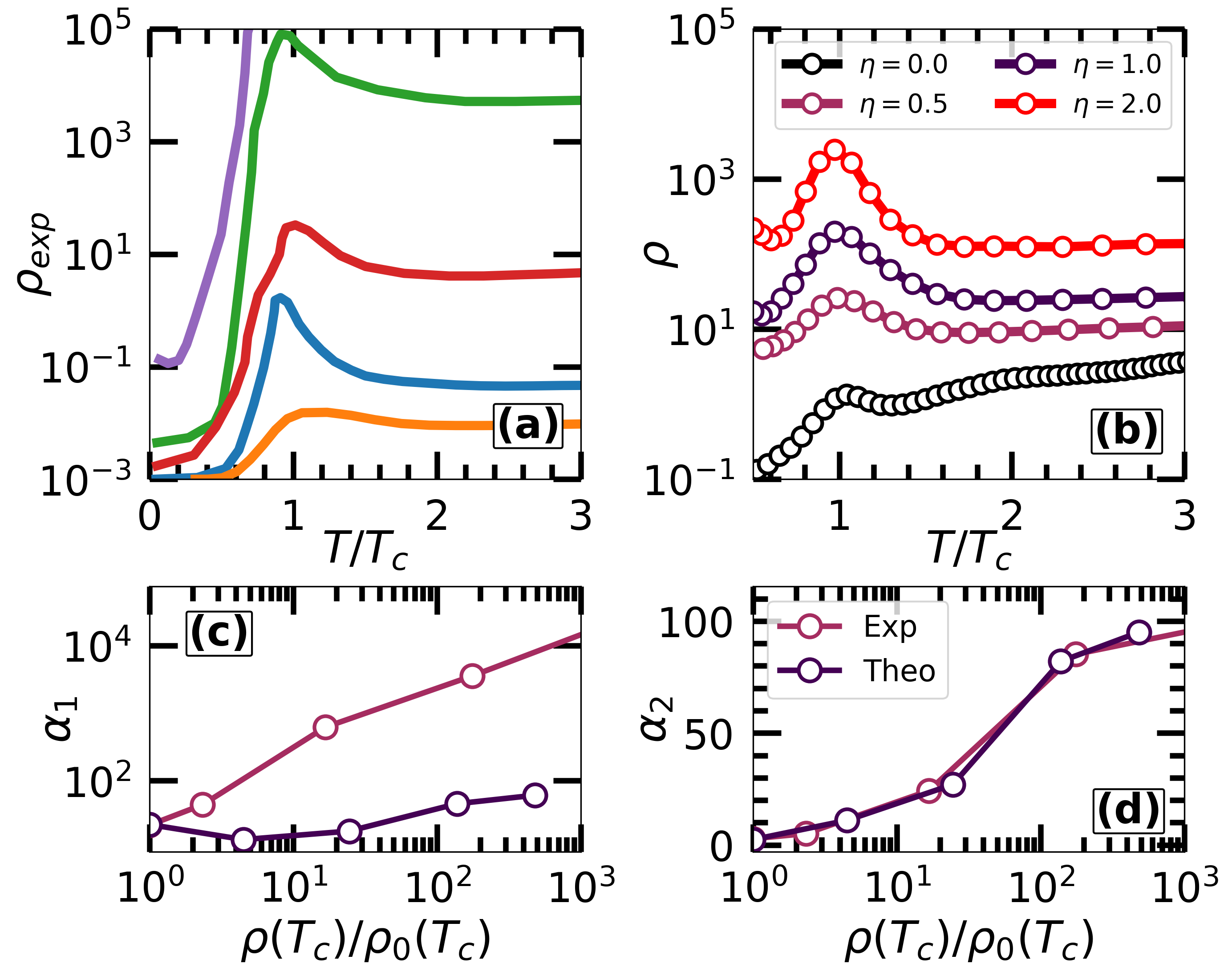}
~~
}
\caption{Results on resistivity.
(a)~Experimental resistivity for various EuO
samples \cite{ref_EuO_res_1,ref_EuO_res_2} showing
 enhanced nonmonotonicity due to impurity effects.
(b)~Our computed resistivity for $n_{el} = 0.025$ and
varying $\eta= n_{imp}/n_{el}$ from $0-2.0$.
Apart from the overall rise with $\eta$, note the clearly
non exponential behaviour at high temperature.
(c)~The ratio $\alpha_1 = \rho(T_c)/\rho(0.5T_c)$ plotted with
respect to $\rho(T_c)/\rho_0(T_c)$, where $\rho_0(T_c)$ corresponds
to the least disordered sample.
While both the experimental and theoretical results increase with
disorder, the match is quite poor. We ascribe this to our
2D setting (see text).
(d)~The ratio $\alpha_2 = \rho(T_c)/\rho(2T_c)$, does far
better in the theory-experiment comparison.
}
\end{figure}
\begin{figure*}[t]
\centerline{
\includegraphics[height=6.5cm,width=17.0cm]{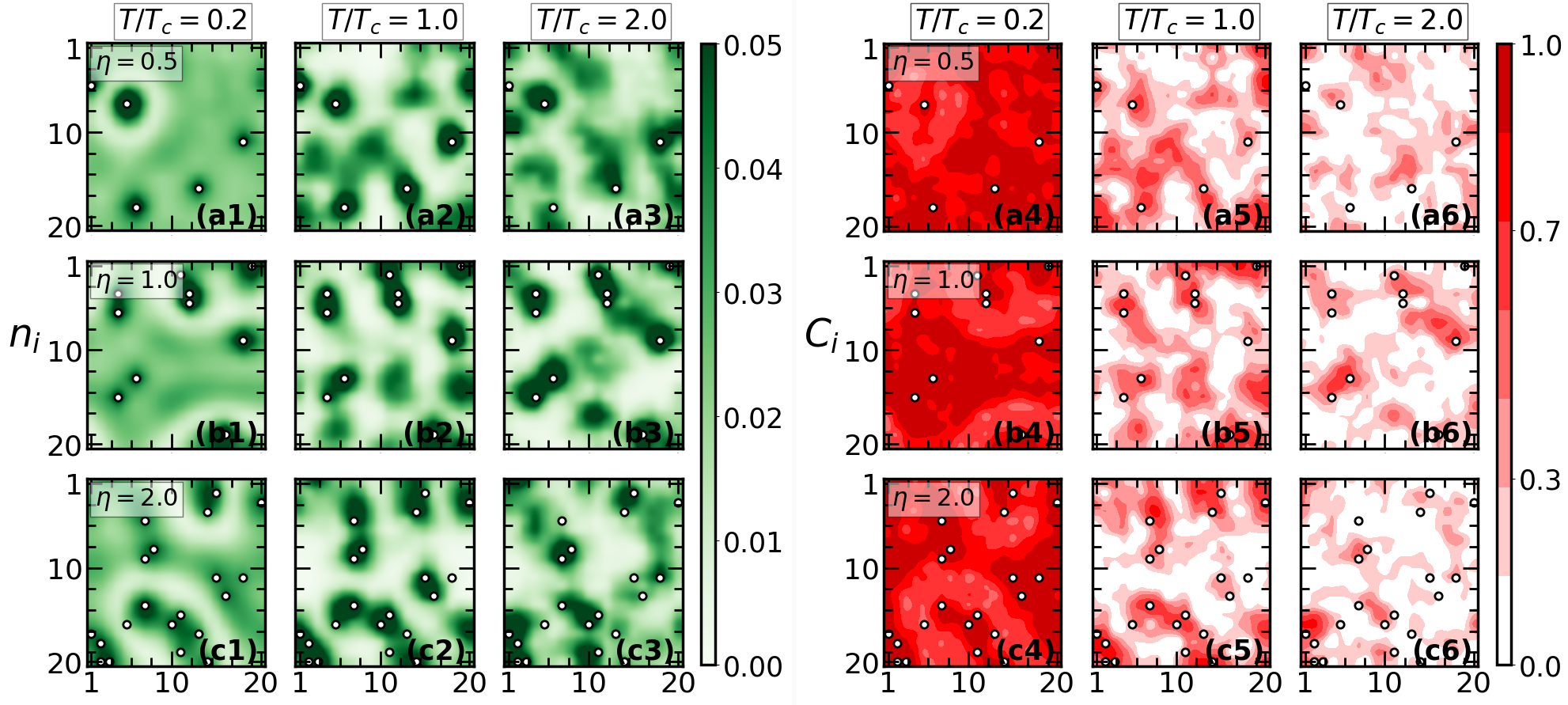}
}
\caption{Spatial signature of disorder enhanced polaron formation. 
Different rows correspond to different  $\eta$.  The left panels
display the spatial map of the carrier density $n_i$, while the
corresponding short range ferromagnetic spin correlation $C_i$ 
(see text) is shown on the right. The data corresponds to $n_{el} 
= 0.025$ and varying $\eta= n_{imp}/n_{el}$ as marked in the panels. 
The impurity locations are marked by small open circles in the 
panels. At all $\eta$ a nearly homogeneous $n_i$ at low $T$ 
becomes increasingly 
inhomogeneous near $T_c$, accompanied by a transition in 
$C_i$ from uniform ferromagnetism to a `phase-separated' pattern, 
signaling FP formation. At higher $T$, the correlation between 
$n_i$ and $C_i$ weakens, indicating the loss of polaronic character. 
Larger $\eta$ enhances inhomogeneity at $T_c$, suggesting stronger 
polaron formation.
} 
\end{figure*}

While some qualitative features, including the nonmonotonicity
and a non exponential high temperature behaviour is common
between the theory and experiment, we attempt a closer
comparison. We do not know the impurity parameters
in the real material so we study two 
`parametric' quantities, with disorder being implicit.
Panels (c) and (d) quantify the peak structure by 
comparing $\rho(T_c)$ to characteristic low and high
temperature values.  We define the ratios,
$\alpha_1 = \rho(T_c)/\rho(0.5T_c)$ and $\alpha_2
= \rho(T_c)/\rho(2T_c)$. We show the comparison
of theory and experiment for $\alpha_1$ and 
$\alpha_2$ in panels (c) and (d), respectively. They 
are plotted against $\rho(T_c)/\rho_0(T_c)$ where 
we use $\rho(T_c)$ as a proxy for the degree of
disorder and $\rho_0$,  the $T=T_c$ 
resistivity of the `cleanest' sample, is
just a normalising scale. 

As panel (c) shows $\alpha_1$ does quite badly in 
capturing the experimental behaviour. The problem
we think is with the 2D treatment of our 
disordered system where, as 
system size $L \rightarrow \infty$, we will always
have a localised state at finite $\eta$. The 
consequence for a finite lattice is a rapid rise
in resistivity at low $T$ with increasing $\eta$, making 
the results unrepresentative of bulk 3D behaviour. 
The higher temperature behaviour, however, where
singular 2D effects are not important, is much
better captured, as the plot for $\alpha_2$ shows in
panel (d). 
 
On the whole, we think that a polaronic framework, and
our specific calculation captures 
the sample to sample variation of EuO resistivity by 
varying only the impurity concentration, except for 
$T \ll T_c$. In particular, the physics in
the $T \gtrsim T_c$ window, with an insulating but 
non exponential resistivity, had remained out of
reach of theories till now.
We discuss a 3D treatment of the low $T$ resistivity
later.

\section{Spatial textures at zero field}

The resistivity suggests an enhancement of 
polaronic effects, which would have been present already
at $\eta=0$, with increasing disorder. To understand 
the impact of disorder on the polarons, we analyze the 
electron density $n_i$ and its relation with local 
magnetic order, as shown in Fig.2. 
Note that we are at the average 
electron density $0.025$ so if the
spatial density had remained uniform, we would be at the
middle of the colour bar, whose peak is set to $0.05$.

\begin{figure}[b]
\centerline{
\includegraphics[height=7.0cm,width=8.5cm]{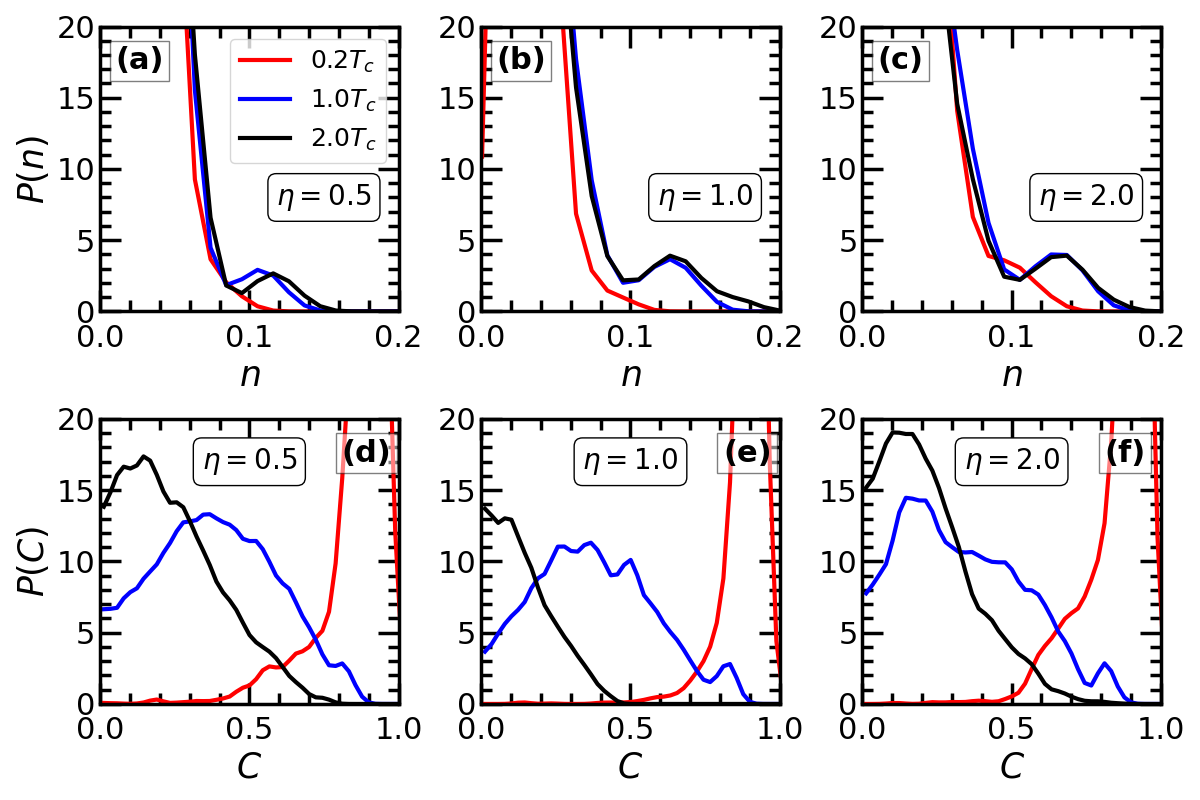}
~~
}
\caption{Probability distribution of density and local
magnetic correlation.
(c)-(c) show $P(n)$ revealing the emergence of a weak
high density feature with increasing $\eta$ at low $T$, 
and a  high density feature which becomes 
progressively prominent with increasing $T$ and $\eta$. 
(d)-(f) show
$P(C)$, the notable feature being the high
correlation peak as $\eta$ increases at $T_c$.
}
\label{fig_histo}
\end{figure}
\begin{figure*}[t]
\includegraphics[height=7.0cm,width=12.0 cm]{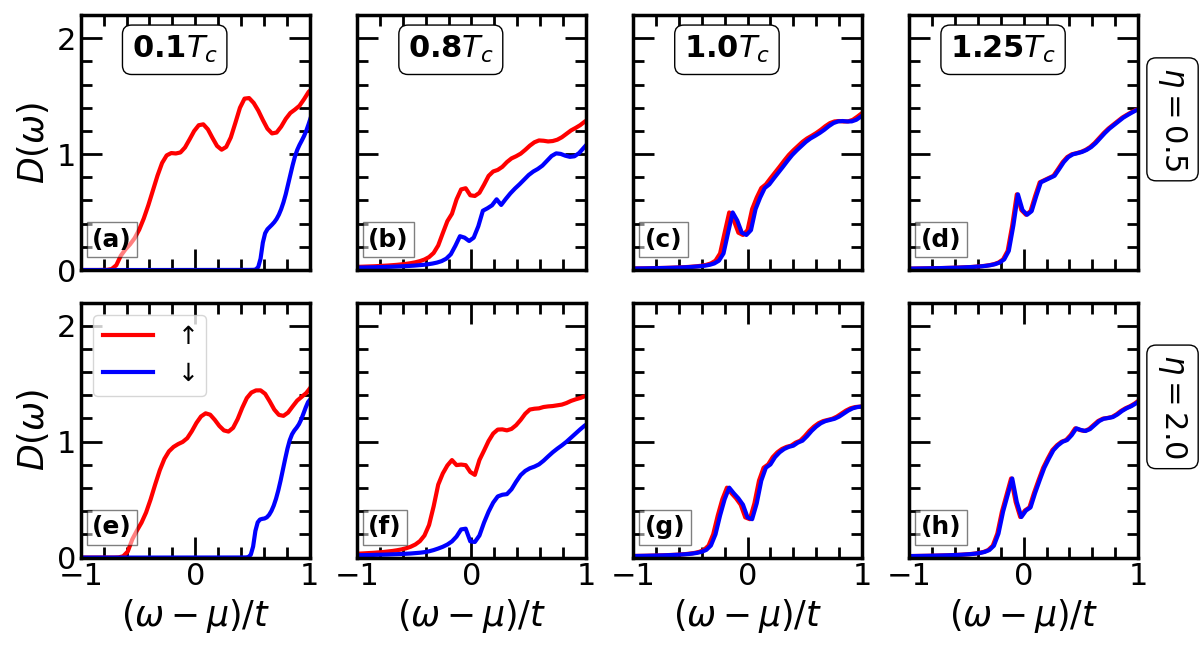}
\caption{Spin-resolved density of states. The top row is 
for $\eta = 0.5$, the bottom for $\eta = 2.0$.
The red line represents the `majority' spin channel,
blue represents the `minority' spin channel.
The feature to note is the emergence of a pseudogap 
at the chemical potential in both channels at $T=0.8T_c$, 
its deepening at $T=T_c$, and weakening at $T=2T_c$. The results 
at weak and strong disorder are similar, except the dip feature 
is more prominent at $\eta = 2.0$.
}
\end{figure*}

Panels (a1)-(a3) show $n_i$ at weak disorder, $\eta=0.5$.
At $T =  0.2T_c$ the electron density is only weakly
inhomogeneous, but at $T = T_c$ the electrons localize, some
around the impurity sites and some elsewhere (there are
more electrons than impurities), evidenced by the appearance 
of dark green patches.  If we examine the magnetic correlations 
in (a4)-(a6) we see that the $T=0.2T_c$ state is almost fully 
polarised while the $T=T_c$ state has residual ferromagnetic
correlations precisely in the regions where $n_i$ is
high. At $T=2T_c$ while there is still a density 
inhomogeneity, the associated spin correlations are much
weaker. The electrons no longer `polarise' their
neighbourhood significantly.
Note that in contrast to the `bound magnetic polaron' scenario
the impurities act as seed for creating an inhomogeneity and
are not the `trapping site' for all the electrons.

The magnetic inhomogeneity 
we observe is consistent with experimental conclusions from $\mu$SR
\cite{ref_EuO_meuon} and Raman scattering
\cite{ref_EuO_raman}. The linear size of the high density
regions is smaller — about 4-5 lattice spacings — compared 
to the disorder-free system \cite{ref_EuB6}, 
indicating stronger electron localization.
At $T = 2T_c$, while density inhomogeneities 
persist the correlation with local magnetic order
weakens.

At  moderate disorder,  $\eta=1.0$, shown in panels 
(b1)-(b6) the density $n_i$ is more inhomogeneous 
even at $T = 0.2T_c$ due to the larger number of
impurities. However, there are no `polarons' at $0.2T_c$ 
as the system maintains nearly perfect ferromagnetic
order.  At  $T =T_c$ the density inhomogeneity is more
pronounced, (b2), and `magnetic phase separation'
emerges, (b5), with ferromagnetic patches surviving
in an overall magnetically disordered environment.
As $n_{el} = n_{imp}$,
the electrons are localised in the vicinity of the impurity
sites. At $T=2T_c$ density inhomogeneity persists, and 
the local ferromagnetic correlations are somewhat 
stronger than at $\eta=0.5$.

Finally, at high impurity concentration, $n_{imp} = 2n_{el}$,
panels (c1)-(c6), the density inhomogeneity at low temperature
is more pronounced as is the pattern near $T_c$ in (c2).
Local ferromagnetic correlations are stronger (which we
quantify in Fig.3). The polarons are more localized,
exhibiting a smaller and more isotropic shape,
the typical linear dimension 
at $T_c$ having reduced from
$\sim 5$ at $\eta=0.5$ to $\sim 3$ at $\eta=2.0$.
At this value of $\eta$, locally well correlated 
ferromagnetic regions become more sharply distinguished 
from the surrounding lattice, which exhibits disordered 
spin orientations. 

Since the spatial maps provide only qualitative information
we constructed the thermally averaged distributions for the
density $n_i$ and the ferromagnetic correlation $C_i$ from the data:
$$
P(n) = {1 \over N} \langle \sum_i \delta(n - n_i) \rangle,
~~~
P(C) = {1 \over N} \langle \sum_i \delta(C - C_i) \rangle
$$

If we focus on panels Fig.3(a)-(c) we see that at low $T$ $P(n)$ 
has a broad trunk at low $n$ with no distinctive large $n$
feature (which would indicate strong localisation). There is
only a mild hint at $\eta = 2.0$. At $T_c$ and $2T_c$, however,
there is a distinct feature at all $\eta$, split away from 
the low $n$ bulk. The location of the peak and its 
height move to larger values with increasing $\eta$ and $T$. Thermal
fluctuations interplay with structural disorder to create polarons.
The typical density associated with polaronic sites increases
from $n \sim 0.1$ at $\eta = 0.5$ to $n \sim 0.14$ at $\eta = 2.0$,
a $\sim 40\%$ change in `area'. 
In panels (d)-(f) the low temperature 
$P(C)$ has the expected peak at $C=1$ (saturated state) 
but the interesting result is at $T = T_c$ where we see 
a stronger large $C$ peak with increasing $\eta$.
At $\eta=0.5$, a mild hump like feature separates  out at 
large $C$, indicating stronger ferromagnetic 
patches in the disordered lattice, which dominates the bulk feature. 
As $\eta$ increases, the feature becomes stronger.

\section{Spectral signature of magnetic polarons}

We have seen that structural disorder 
enhances magnetic polaron formation through stronger localization. 
In this section, we investigate the impact of disorder-induced 
 polaron formation on the density of states (DOS).
 Fig.4 shows the majority (red) and minority 
 (blue) DOS for weak and strong disorder scenarios
 in the top and bottom panels, respectively.
The `majority' and `minority' are decided by projecting
electron spin parallel and antiparallel to the magnetisation.
Obviously the distinction does not exist for $T \ge T_c$.

The enhanced polaronic behaviour manifests prominently 
near $T_c$ as a pseudogap in the density of states (DOS) at 
the Fermi level. Unlike in clean systems, a well-defined 
 pseudogap appears already at $\eta=0.5$. 
 This arises as electrons localize to form FP, leading to the 
 formation of bound states at the lower edge of the DOS. 
The pseudogap is deeper and broader at $\eta = 2.0$.

As we move to lower $T$ the pseudogap weakens.
At $0.8T_c$, the growth of ferromagnetically correlated 
regions allows electronic states to delocalize, diminishing
the gap. The pseudogap is still visible in the strong 
disorder case, but nearly vanishes for 
$\eta = 0.5$. For $T \lesssim 0.5T_c$, 
there is no pseudogap at either $\eta$.
As we move above $T_c$,
thermal fluctuations weaken magnetic self trapping, and as 
a result the density inhomogeneity.
As a consequence, the pseudogap
weakens as well, although remnants are visible at
$T=1.25T_c$ for both $\eta = 0.5$ and $2.0$.

\begin{figure}[b]
\centerline{
\includegraphics[height=6.7cm,width=8.5cm]{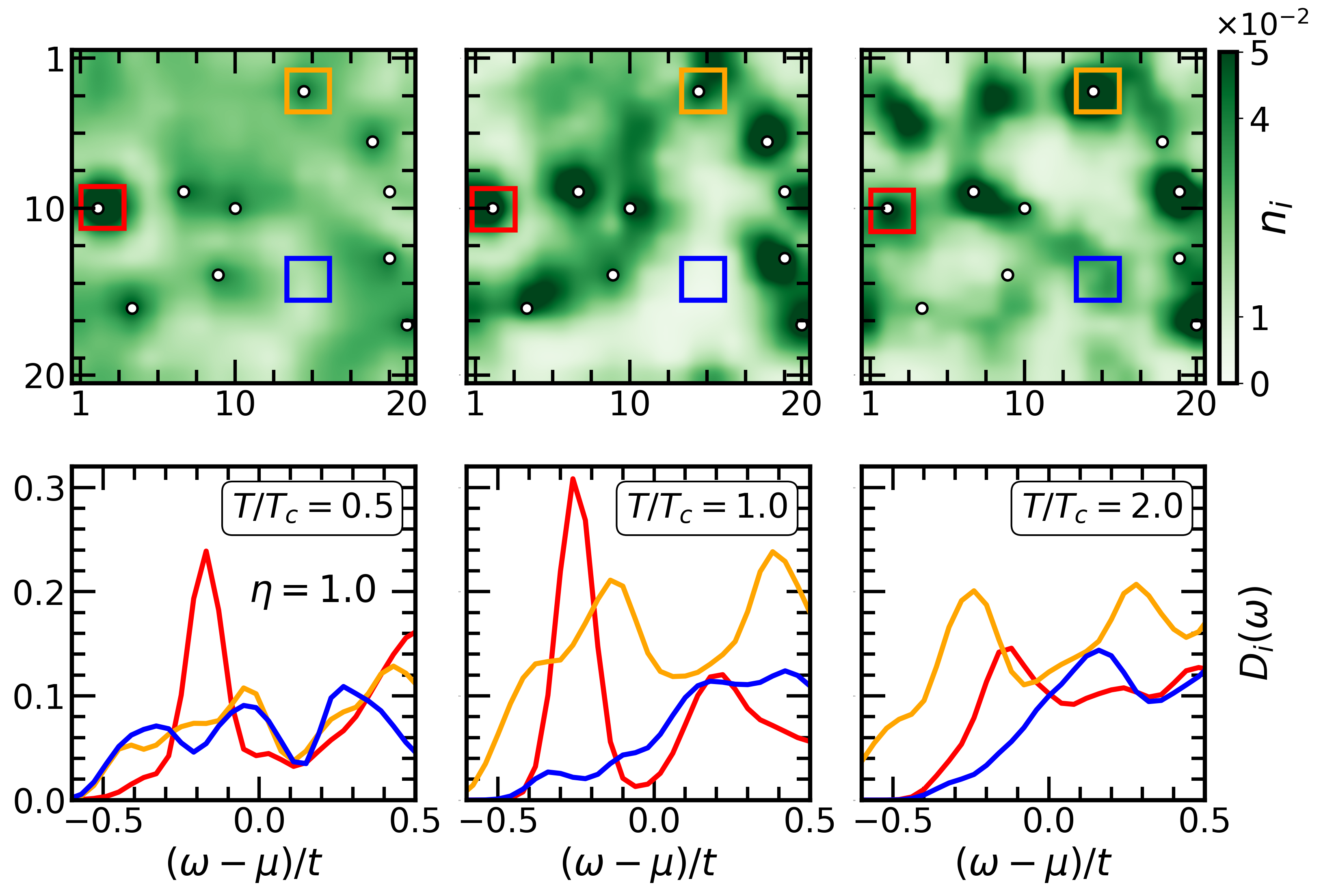}
~~ }
\caption{LDOS at three sites in the system at $\eta=1.0$.
In the upper panels, 
the blue box represents a lattice site without any
impurity potential, while the red and orange boxes are 
around impurity sites. While there is no prominent difference
in the LDOS between the three sites at $0.5T_c$, at $T_c$ the `red site' shows a huge dip at $\mu$ and the `orange site' has a smaller dip. At $2T_c$ the difference between the 
LDOS at the three sites is again small.
}
\end{figure}
\begin{figure}[t]
\centerline{
\includegraphics[height=8.2cm,width=8.0cm]{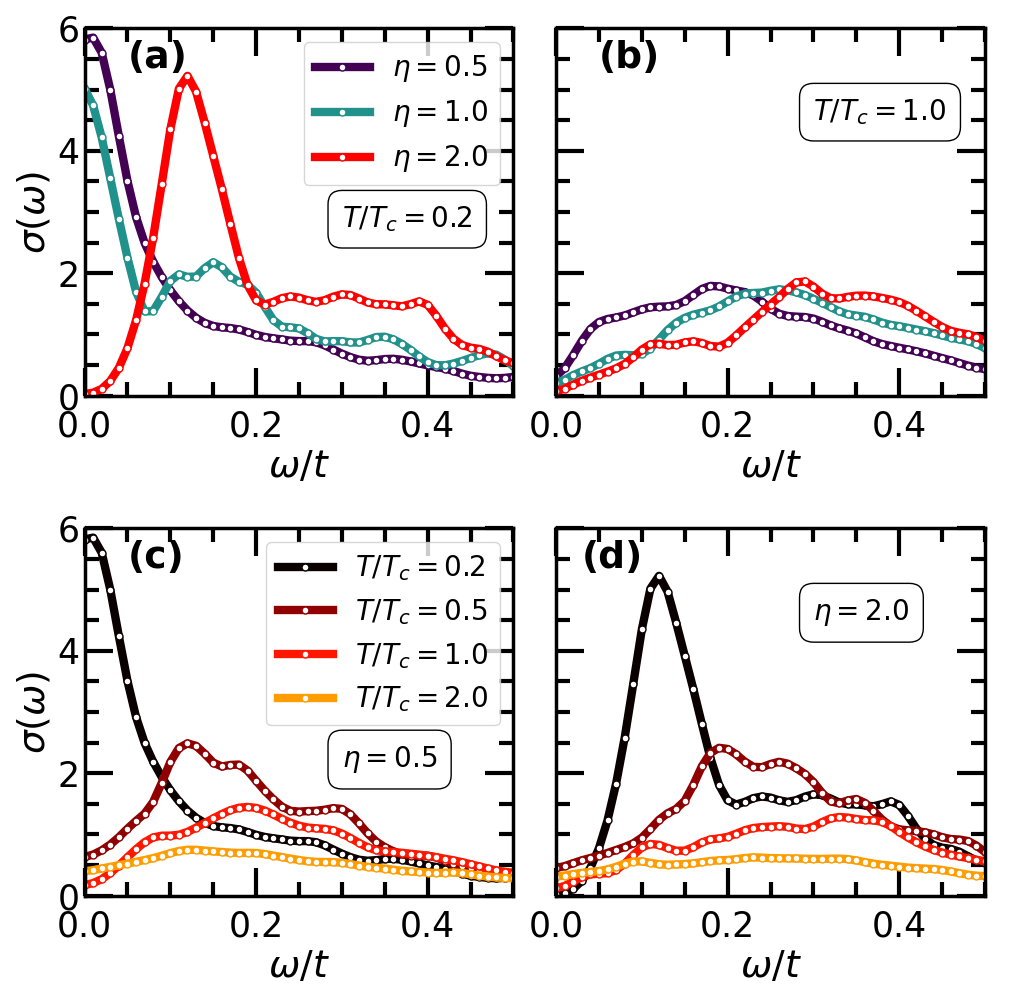}
~~
}
\caption{Optical conductivity: (a)-(b) show the disorder dependence
of $\sigma(\omega)$ at $T=0.2T_c$ and $T=T_c$, respectively.
In (a) we observe the shift from a Drude response at $\eta = 0.5$
to a distinctly non Drude, finite frequency peak, in $\sigma(\omega)$
at $\eta=2.0$. In (b) the $\sigma(\omega)$ is non Drude and broad
at all $\eta$ with the peak shifting to larger frequency with $\eta$.
(c)-(d) Show the $T$ dependence at two $\eta$. At $\eta = 0.5$ there
is a thermal transition from Drude to non Drude behaviour while
at $\eta=2.0$ the response is non Drude at all temperatures.
}
\label{fig_opt}
\end{figure}

The local density of states (LDOS) further confirms the
presence of a gap-like feature around impurity sites. In
Fig.5, we plot the LDOS averaged over a 
$3 \times 3$ region around two impurity sites (red and 
orange) and a non-impurity site (blue). Thermal
averaging is done over 200 configurations, with
a separation time of $5t^{-1}$ in each. The key observation
is that a clear gap-like 
feature exists in the LDOS at both impurity sites
whereas it is absent at the non-impurity 
site. This indicates that impurities act as pinning 
centres for polarons. Even after a total averaging
time of $1000t^{-1}$, the gap-like feature remains visible
at the impurity sites.
We present this as a prediction that can be verified 
experimentally by measuring tunnelling conductance 
as done in the case of EuB$_6$ \cite{ref_EuB6_stm}.

\section{Optical Conductivity}

In Fig.6 we show the optical conductivity $\sigma(\omega)$. 
Panels (a)-(b) show the disorder dependence at two temperatures.
Panel (a) shows the $\eta$ dependence at $T = 0.2T_c$.
For $\eta = 0.5$, a clear Drude-like response is observed, but
with increasing disorder, the low-frequency spectral weight 
diminishes, and at $\eta = 1.0$ a secondary peak emerges 
around $\omega \sim 0.2$, accompanied by a suppression of the 
$\omega = 0$ component. For $\eta = 2.0$, the $\omega = 0$ peak 
vanishes entirely, and the finite-frequency peak at $\omega \sim 0.2t$ 
becomes dominant. This shift from Drude like behaviour to a 
finite energy peak is due to incipient localisation of the 
electron with an increase in disorder, not because of `polaron' 
formation. 
In panel (b), where $T=T_c$, both impurity effects and magnetic
fluctuations are at play, and the self trapped electrons have
no Drude response at any $\eta$. 
$\sigma(\omega)$ is broad, with a peak that shifts outward from
$\omega \sim 0.2t$ as disorder increases beyond $\eta = 0.5$.

Panels (c)-(d) focus on temeperature dependence at fixed
disorder.
At $\eta=0.5$, panel (c), the system exhibits a Drude like
response characterized by a sharp peak at $\omega = 0$ at 
low temperatures ($T = 0.2T_c$). As the temperature increases 
to $0.5T_c$, this peak shifts to a finite frequency,
leading to a suppression of the d.c. conductivity. 
This deviation from Drude behaviour becomes
more pronounced at higher temperatures.
In contrast, at $\eta=2.0$, shown in panel (d), 
we see a qualitatively different behaviour. Already
at $T = 0.2T_c$ no Drude-like peak is 
observed. Instead, a finite-frequency peak emerges 
due to the localization of carriers around impurity 
sites.  As temperature increases, this 
peak shifts to higher frequencies, accompanied by
a significant suppression of d.c. conductivity compared 
to the weak disorder case. This stronger suppression 
aligns with the enhanced non-monotonicity of $\rho(T)$.

\begin{figure}[b]
\centerline{
\includegraphics[height=8.4cm,width=8.5cm]{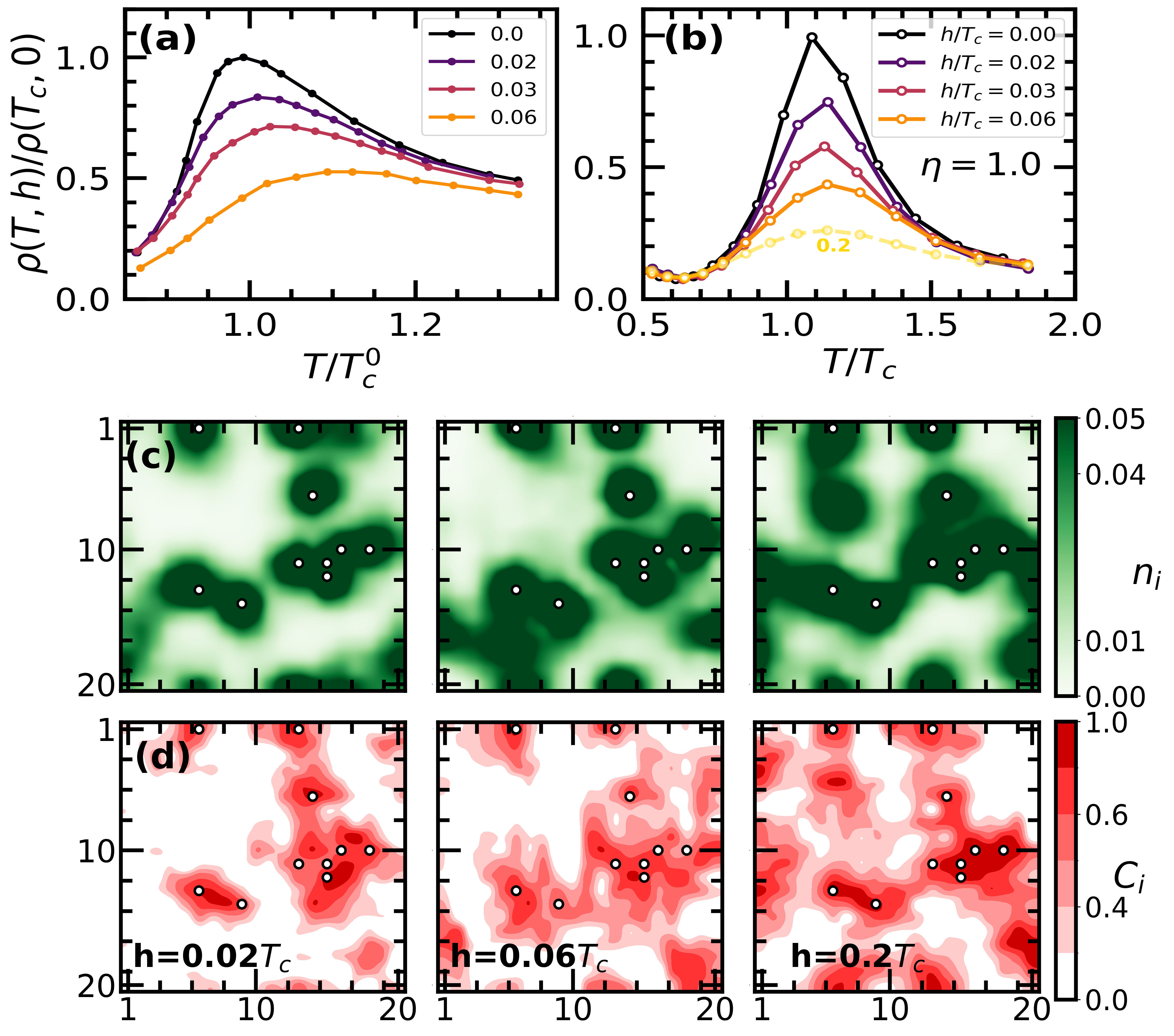}
~~
}
\caption{Magnetic field response: (a) and (b) show the field 
dependence of $\rho(T)$, respectively,  in the experiment 
and within our calculation. Applied fields are calibrated 
as fraction of $T_c$.  (c)~Spatial map of electron density 
$n_i$ at $T_c$ for $h/T_c = 0.02$, $0.06$, and $0.2$. Note 
the increasing spread of the density field, even though some 
inhomogeneity survives at the highest field used. (d)~Local 
ferromagnetic correlation $C_i$ of the spins corresponding 
to (c), showing the inhomogeneous increase in the local 
magnetisation.
}
\end{figure}
\begin{figure}[t]
\centerline{
\includegraphics[height=4.4cm,width=7.5cm]{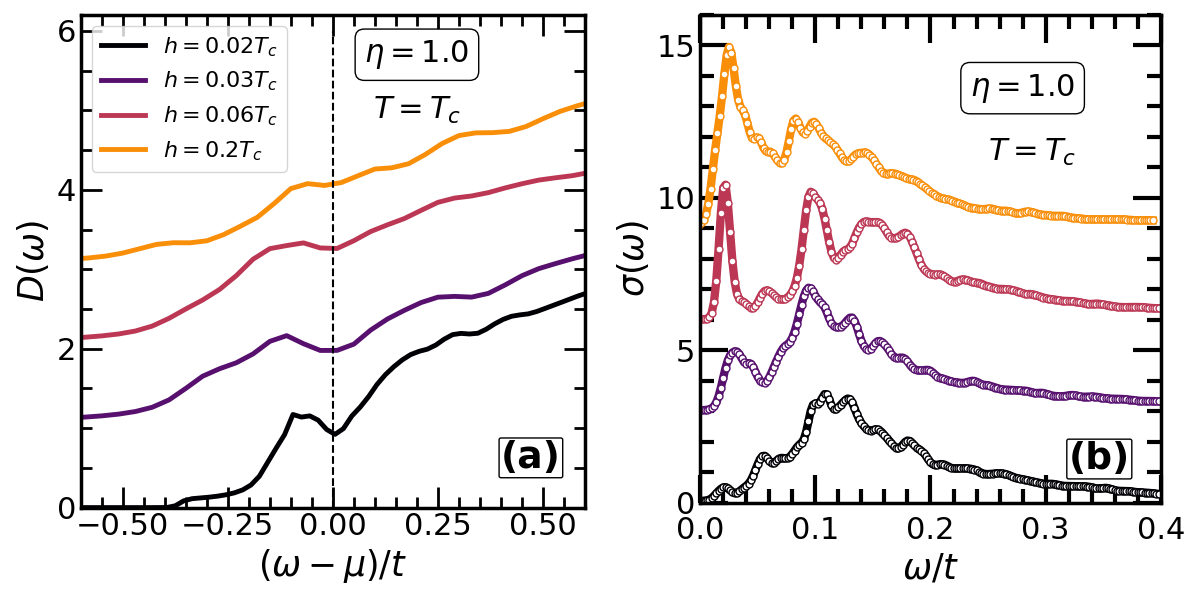}
~~
}
\caption{Effect of an applied field at $T=T_c$ on the
pseudogap and the optical conductivity.
(a)~Density of states 
$D(\omega)$ at $T=T_c$ and $\eta=1.0$ for $h/T_c=
0.02$, $0.03$, $0.06$, and $0.2$. The pseudogap
in the DOS reduces progressively with the increase in $h$ 
and vanishes completely for $h/T_c=0.2$. 
(b)~Optical conductivity $\sigma(\omega)$ for the same 
parameter values shows a gradual transition from a peak feature
at $\omega \sim 0.12t$ at $h=0$ to an `almost Drude like'
response at $h/T_c = 0.2$ with the peak pushed down to $\omega
\sim 0.02t$. For clarity, the plots in both panels are vertically shifted relative to the adjacent ones below by $17\%$ and $20\%$ of their respective maximum $y$-ranges for panels (a) and (b), respectively.}
\end{figure}

\section{Magnetoresistance and homogenisation}

We now examine the effect of an external magnetic field on
the nonmonotonic temperature dependence of $\rho(T, h)$. 
Focusing on the case $\eta  = 1.0$, we vary
$h$ in the range $0–0.2T_c$ and employ a field-cooling
protocol to obtain equilibrium spin configurations at each
temperature. 

As shown in Fig.7, the application
of a magnetic field leads to a pronounced suppression of
the resistivity peak at $T_c$.
This behaviour qualitatively resembles
the experimental magnetoresistance reported for
sample 4B in Ref.\cite{ref_EuO_res_1}. Notably, due
to the large initial ratio $\rho(T_c,0)/\rho(2T_c,0)$,
a weak nonmonotonic feature persists 
even at $h = 0.07T_c$.

In panels (a)-(b) we show, respectively, 
the field dependent resistivity in
an EuO sample, sample $4B$ from Ref.\cite{ref_EuO_res_2},
and that computed within our theory at $\eta=1$.
Both plots are normalised by the respective $h=0$ peak
values so the maximum is $1$.
The experimental sample 
exhibits a $\rho(T_c)/\rho(2T_c)$ ratio in the
same ballpark as our $\eta=1$ sample. 
We express $h$, both in experiment and theory, as a fraction of
the $T_c$ scale - rather than in Tesla or a fraction of 
hopping - since the $h/T_c$ ratio naturally captures the
effect of field induced polarisation versus thermal fluctuation.
We use the dimensionless magnetic field
$h \rightarrow (g \mu_B S /{k_B T_c})h$,
 where $g$ is the gyromagnetic ratio, $\mu_B$ is the Bohr 
 magneton, $S$ is the local moment of EuO, and $k_B$ is the 
 Boltzmann constant. 

In panel (a) as $h$ increases the non-monotonicity in 
$\rho(T)$ near $T_c$ is progressively suppressed, with 
a weak signature remaining at $h = 0.06T_c$. The 
theoretical result in panel (b) has a $h=0$ peak 
structure that is sharper so the sample is `more 
insulating' compared to the experimental case.
Nevertheless, the fractional drop in resistivity at
low field and $T\sim T_c$ are comparable. Even better 
correspondence can be obtained by studying a weaker 
disorder case theoretically, which we are doing. On 
the whole there is a roughly $30\% $ drop at $h/T_c 
= 0.02$ and a $\sim 40\%$ drop at $h/T_c = 0.03$ 
in both experiment and theory.

The experiments do not reveal what might be happening spatially, 
and here theory can add some insight. In row (c) analyze the 
density $n_i$ and in (d) we examine the local spin correlation 
$C_i$ at $T_c$ for field strengths $h/T_c = 0.02, 0.06$, and 
$0.2$.  $n_i$, which is concentrated at impurity sites at $h=0$, 
spreads out significantly by the time $h/T_c = 0.2$, creating a
percolating pattern.  As Fig.11 will later show the area 
associated with the electronic wavefunctions near the chemical 
potential expands by a factor of $4$ at $h/T_c = 0.2$ with 
respect to the $h=0$ case. $C_i$ shows a strong field 
dependence. As $h$ increases, ferromagnetic correlations are 
enhanced, leading to the emergence of broader ferromagnetic 
patches.

What about signatures in the single particle and optical
spectrum? 
In Fig.8(a) we examine the evolution of the PG in the DOS. 
The PG, present at $T_c$, 
gradually diminishes with increasing $h$ and eventually 
disappears at $h/T_c = 0.2$.
A consistent effect is observed in the optical conductivity 
shown in panel (b), where increases in $h$, cause a gradual
transition from a finite $\omega$ peak 
to a (quasi) Drude behaviour at large field. 

\section{Discussion}

We have presented results on various physical quantities and 
suggested what might be happening in EuO. We feel we also need 
to (i)~create a qualitative picture beyond the numerical results,
particularly focusing on the electronic eigenstates in the self-trapped
regime, (ii)~confirm that our parameter variation covers the
experimental situation, and (iii)~comment on how the resistivity 
trends could change in three dimensions, where weak localisation 
effects, etc, are absent. 

\subsection{Interpreting resistivity: character of eigenstates}

For a given set of impurity positions, quantities like 
the conductivity, density of states, etc, are computed over 
individual spin configurations 
and then thermally averaged.  The key inputs are the eigenstates
$\psi_n^{\alpha}(i)$ and $\epsilon_n^{\alpha}$, where $\alpha$ 
is a spin configuration index. Since it is difficult to visualise 
a large number of eigenstates, we calculate the inverse participation 
ratio (IPR), $I_n^{\alpha}$, of a state and use its inverse, 
$A_n^{\alpha}$, as a measure of the `area' covered by the eigenstate. 
For a plane wave state in 2D, $I = 1/L^2$ and $A = L^2$. We can 
look at $A$ as a function of the energy $\epsilon_n$. D.c transport
at low $T$ will be governed by current matrix elements between pairs 
of states, one on each side of the chemical potential.
 
Before looking at the actual data on $A(\epsilon_n)$ for typical
spin
configurations at different $T$, we try to build a scenario.
We assume that we are at a coupling $J'$ where a magnetic 
polaron can form near $T_c$ in a structurally 
non disordered system. The physics of this 
is known \cite{ref_theory_1polaron},
and we take it as our starting point. We want to consider 
the effect of a `moderate' impurity potential, like $V=-2t$, and 
low but finite electron density ($N_{el} \ll N$) and speculate what
might happen as $N_{imp}$ grows from small values and increases 
beyond  $N_{el}$. Our arguments are not specific to 2D, and 
in fact should work better in 3D where weak localisation effects, 
etc, are absent.

First, the one impurity one electron problem.
At low temperature, where the system is almost fully magnetised
the electron would sense only the impurity potential, leading
to a weak inhomogeneity in $n_i$. Even for the lowest  states 
the `area' would be $O(L^2)$. As $T$  increases, the effective 
potential seen by the electron picks up a magnetic component, 
to be self consistently determined. We can directly estimate 
this from available numerical data \cite{ref_theory_1polaron}  
or write an approximate effective potential of the form:
$ V_{eff} = V_{imp} + V_{mag}(T,l_p) $, where $l_p$ is a 
variational length associated with the polaron size.
Adding the electron hopping term and minimising the total
energy
with respect to $l_p$ leads to a lengthscale $L_p(V,T)$. It is 
harder but numerically possible to work out the full spectrum 
$D(\omega)$. One expects solutions with  $L_p \ll L$ above some 
characteristic temperature $T_p(V)$. In a many electron system,
we would want $L_p \lesssim L/\sqrt{N_{el}}$. 
For $T_p < T \lesssim T_c$, the size  $L_p$ would decrease  
with increasing $T$.  Now, the many 
electron, many impurity, case.

(i)~$N_{imp} \ll  N_{el}$: For $T < T_p$ there would be no
polarons, and one would have a weakly inhomogeneous metallic
state. Now consider the $T > T_p$ situation, and assume that 
the mean inter-impurity separation is $\gg L_p$, so that the 
single (or isolated) polaron results have some use.
The impurities will serve as localisation centers
for only $N_{imp}$ electrons and $N_{el} - N_{imp}$ electrons 
would be delocalised (or weakly trapped). 
So, the `occupied states' in the $N_{el}$ system
will consist of low lying bound states {\it and higher energy 
scattered states}.  $A_n$ will vary from small values to 
$O(L^2)$ even for $\epsilon_n \le \mu$, and would of course
be $O(L^2)$ for $\epsilon_n > \mu$.

The current matrix elements, that dictate the conductivity,
would be between the delocalised occupied states below $\mu$ 
and the delocalised empty states above $\mu$. One would get a 
``bad'' metal, partly due to the strong scattering of the 
delocalised states and the small $n_{el}$ (and associated small
Fermi surface). As $T$ grows, the increasing magnetic scattering
of the delocalised states would lead to larger resistivity, but 
the deeper `polaronic' states do not have a role in the d.c
conductivity! What happens as $N_{imp}$ grows?

\begin{figure*}[t]
\centerline{
\includegraphics[height=4.5cm,width=13.5cm]{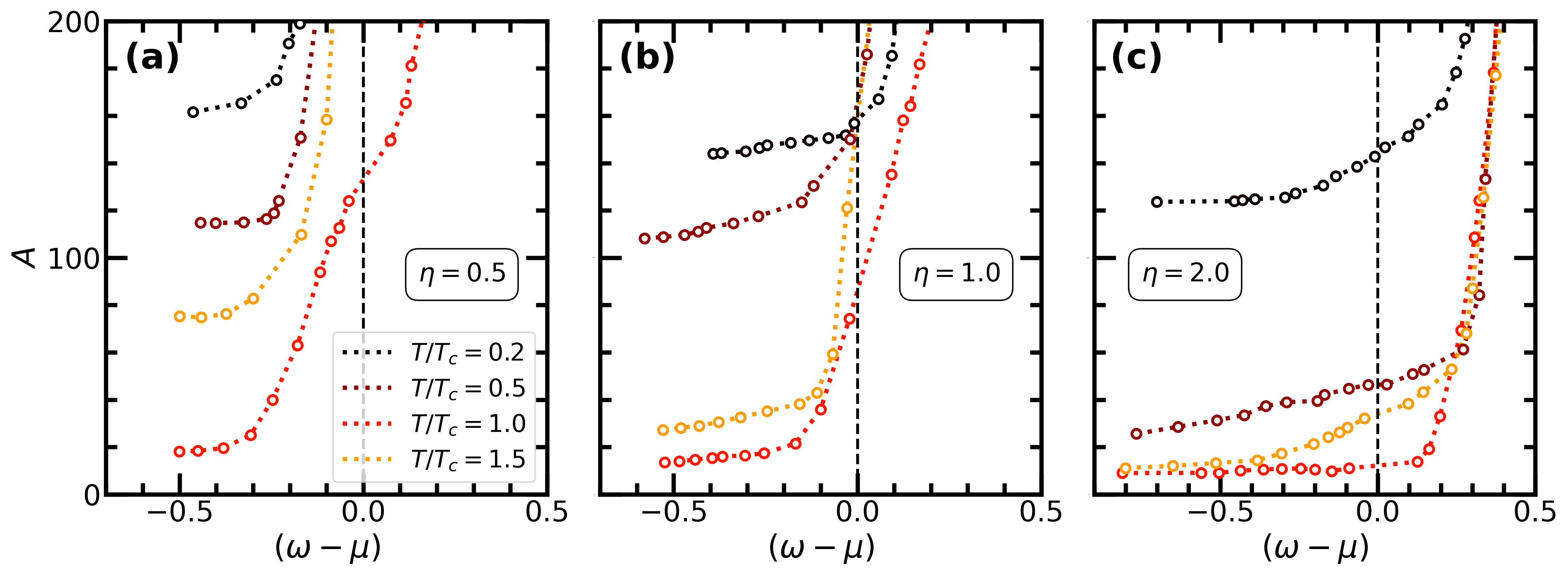}
~~
}
\caption{The area $A$, computed as the inverse of the inverse 
participation ratio (IPR) for typical equilibrium spin configurations. 
Since $N=L^2 = 400$ and $N_{el} =10$ we use $A \le 40$ to identify 
localised states.  (a)~At $\eta = 0.5$ there is no localisation at 
$T= 0.2T_c$ and $0.5T_c$. At $T=T_c$ (bottom curve), about $50\%$ of 
the occupied states are localised, and states at $\mu$ are delocalised, 
and at $T=1.5T_c$, nothing is localised. (b)~At $\eta=1.0$, nothing
is localised at the lowest two temperatures. At $T_c$ (bottom
curve) $90\%$ of the states below $\mu$ are localised. At
$1.5T_c$, that number is $\sim 60\%$, and the area associated
with the localised states is also larger. 
(c)~At $\eta=2.0$  all states are delocalised at $0.2T_c$ but
about $60\%$ of the occupied states are localised at $0.5T_c$.
At $T_c$ and $1.5T_c$ not only are all states below $\mu$
localised, some states above $\mu$ are localised as well. 
There is a remarkably quick rise in $A$ beyond a small energy
above $\mu$.}
\end{figure*}

(ii)~$N_{imp} \sim N_{el}$:
Here, again, the low $T$ picture is similar to the
earlier case, but after polaron formation with increasing
$T$ all the electrons can localise around impurity sites.
So, for $\epsilon_n \le \mu$ we should have
$A_n \ll L^2$ and for $\epsilon_n > \mu$ the
area $A_n$ should rapidly grow to $O(L^2)$.
The conductivity  will be controlled now by matrix elements
between bound states below $\mu$  and extended or (weakly bound)
states above $\mu$.
As the polaronic state becomes more compact with increasing
$T$  or $V$ the matrix elements reduce, suppressing
the conductivity.

(iii)~$N_{imp} \gg  N_{el}$:
Skipping to the $T > T_p$ scenario,
in this case, there are potentially more localisation sites
available than the number of electrons. While $N_{el}$
impurity sites can actually host polaronic states, there are $N_{imp}
- N_{el}$ additional sites with potential $-V$. While there
are no non trivial magnetic correlations around these, the
combination of impurity potential and magnetic disorder can
create bound states above $\mu$ as well. For $\epsilon_n \gg \mu$
of course we should have $A_n \sim L^2$, but the d.c conductivity
would now arise from matrix elements between localised states
below $\mu$ and localised states above $\mu$.
In these self consistently computed equilibrium configurations
we could have a `mobility edge' somewhere above $\mu$ so
that conductivity depends on activation to a higher energy
delocalised states.

What does the numerics show? Fig.9 shows $A$ as a function of 
energy of the eigenstates, for single typical configurations, 
at four temperatures for three values of $N_{imp}/N_{el}$. The 
vertical dotted line sets the chemical potential in each panel.
We have checked that choosing some other equilibrium spin
configuration does not change the qualitative features of the
result.

Panel (a) shows results at $\eta = 0.5$. If we set $A < L^2/N_{el}
= 400/10$ as a cutoff for the existence of a polaron, there are no 
polaronic states for $T=0.2T_c$ and $0.5T_c$. At $T_c$
about $50\%$ of the occupied states meet this criterion
while at $1.5T_c$ there are again no polaronic states. So (i)~a
very narrow window around $T_c$ where polarons exist, and (
ii)~the d.c. conductivity will be decided by delocalised, albeit
strongly scattered, states. 

Panel (b) shows results at $\eta = 1.0$. Again, no localisation
at the lowest two temperatures but quite low area, $A \sim 10$
for the low lying occupied states at $T=T_c$. At $T_c$ about
$90\%$ of the occupied states are localised, at $1.5T_c$
this fraction is $\sim 60\%$ (and the associated $A$ are
larger). Polarons form below $T_c$ and survive to $1.5T_c$.
The low $T$ conductivity would be dictated by pairs of
delocalised states on either side of $\mu$ while at $T_c$
it would involve matrix elements between a localised and a 
delocalised state. The results both at $\eta = 0.5$ and
$\eta=1.0$ bring to life the fact that even when polarons
exist in a system {\it not all electrons need to be in 
polaronic states}. Solving the one polaron problem 
does not solve the many polaron problem except in 
the extreme dilute limit.

Panel (c) shows results at $\eta=2.0$. While $T=0.2T_c$
expectedly does not have localised states, about
$60\%$ of the occupied states are localised at $T=0.5T_c$
(the fraction would be $\sim 100\%$ if we had a more
generous cutoff). At $T_c$ and $1.5T_c$, all occupied
states are localised and indeed some states above
$\mu $ as well. Bulk conduction would have to arise
from activation to delocalised states which are at
some energy $\sim 0.2t-0.3t$ above $\mu$. 

On the whole, the explicit results bear out our basic
suggestion about the changing fraction of localised
states as a function of $N_{imp}$ and $T$, and that
the conduction would be decided by different kinds
of pairs of states in the three $\eta$ regimes.

\begin{figure}[b]
\centerline{
\includegraphics[height=5.5cm,width=8.5cm]{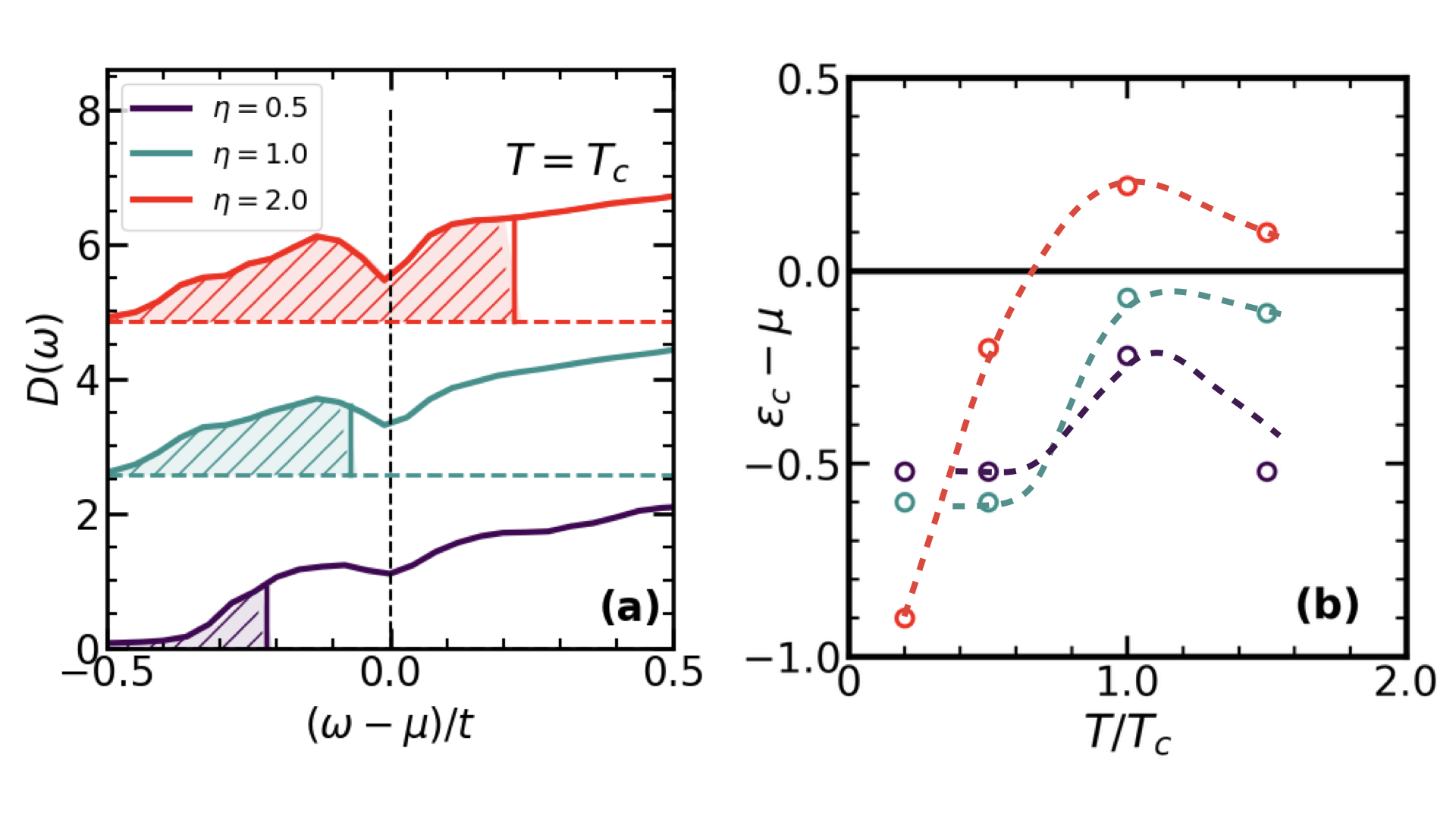}
~~
}
\caption{Localised states and mobility edge: (a)~Shows
the density of states at $T_c$ for three values of $\eta$,
with the shaded area representing the localised states. As
$\eta$ grows from $0.5$ the mobility edge moves from below
$\mu$ to near $\mu$ and then, at $\eta = 2.0$ to clearly
above $\mu$. For clarity, the plots are vertically shifted relative to the adjacent ones below by $30\%$ of the maximum $y$-ranges. (b)~A crude `trajectory' of,
 $\epsilon_c - \mu$, the mobility edge with respect to
the chemical potential, with changing $T$. 
}
\end{figure}

Since much of the discussion is about the fraction of localised 
states in the band, one may wonder if we can locate a `mobility 
edge', $\epsilon_c$, as one does in usual disordered systems, 
as a function of $\eta$ and $T$. While the effective `disorder' 
here is annealed - it depends on the solution of a complex thermal 
problem - and is not externally specified, locating  $\epsilon_c$ 
can create some insight.
To that extent Fig.10(a) shows the density of states at $T=T_c$
for the three $\eta$ values and identifies the localised states
by shading. The upper limit of the shaded area is $\epsilon_c$.
Starting at the bottom, at $\eta = 0.5$, only a fraction of the
occupied states is localised and $\epsilon_c < \mu$. At $\eta 
= 1.0$ almost all occupied states are localised and $\epsilon_c 
\approx \mu$,
while at $\eta = 2.0$ localised states extend beyond $\mu$ and
we have $\epsilon_c > \mu$. While the background configurations
emerge from a difficult annealing problem, once obtained it is
pictures like panel (a) that help understand transport. Note 
also the deepening pseudogap around $\mu$ as $\eta$ increases,
already seen in Fig.4

\begin{figure}[b]
\centerline{
\includegraphics[height=5.8cm,width=8.5cm]{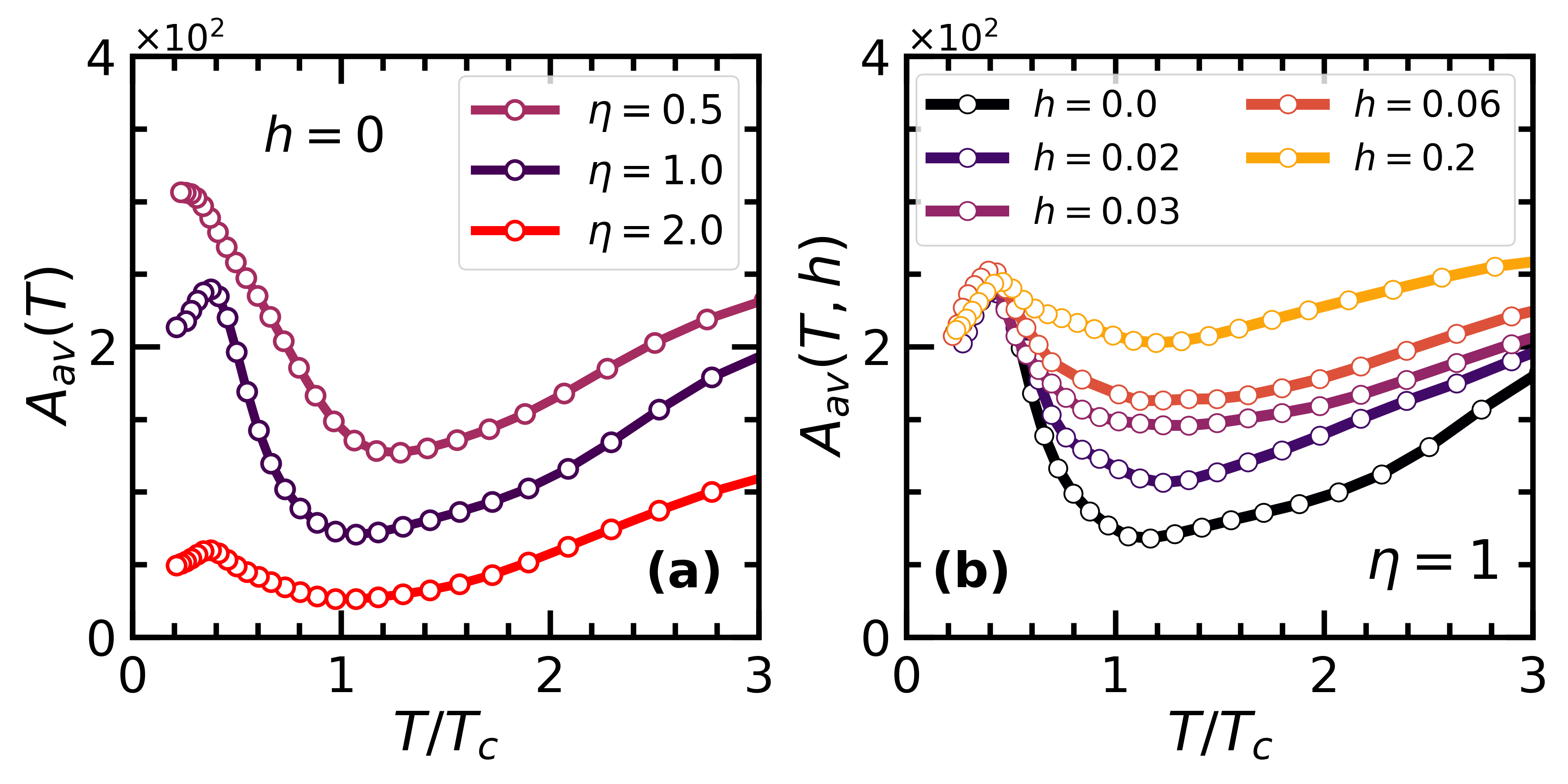} ~~ }
\caption{Average area $A_{av}$ associated with the eigenstates
that lie in the energy window $\mu \pm T$.  (a)~$A_{av}$  at
$h=0$ for $\eta=0.5$, $1.0$, and $2.0$. It shows relatively
large values at $\eta=0.5,~1.0$ at low $T$ due to the
delocalised nature of states, followed by a sharp dip near
$T_c$. At $\eta=2.0$, the low $T$ value is already small, and
the bound states become more compact as $T \rightarrow T_c$.
As $T$ increases beyond $T_c$, the average area rises again.
The non monotonicity reflects in the non monotonic $\rho(T)$.
(b)~$A_{av}$ at $\eta=1.0$ for $h/T_c = 0$,
$0.02$, $0.03$, $0.06$, and $0.2$. Here we see the field
driven increase in spatial coverage, and almost
`delocalisation' at large $h$.
A plane wave state on our lattice would have $A = 400$.
} 
\end{figure}
\begin{figure}[t]
\centerline{
\includegraphics[height=5.2cm,width=8.5cm]{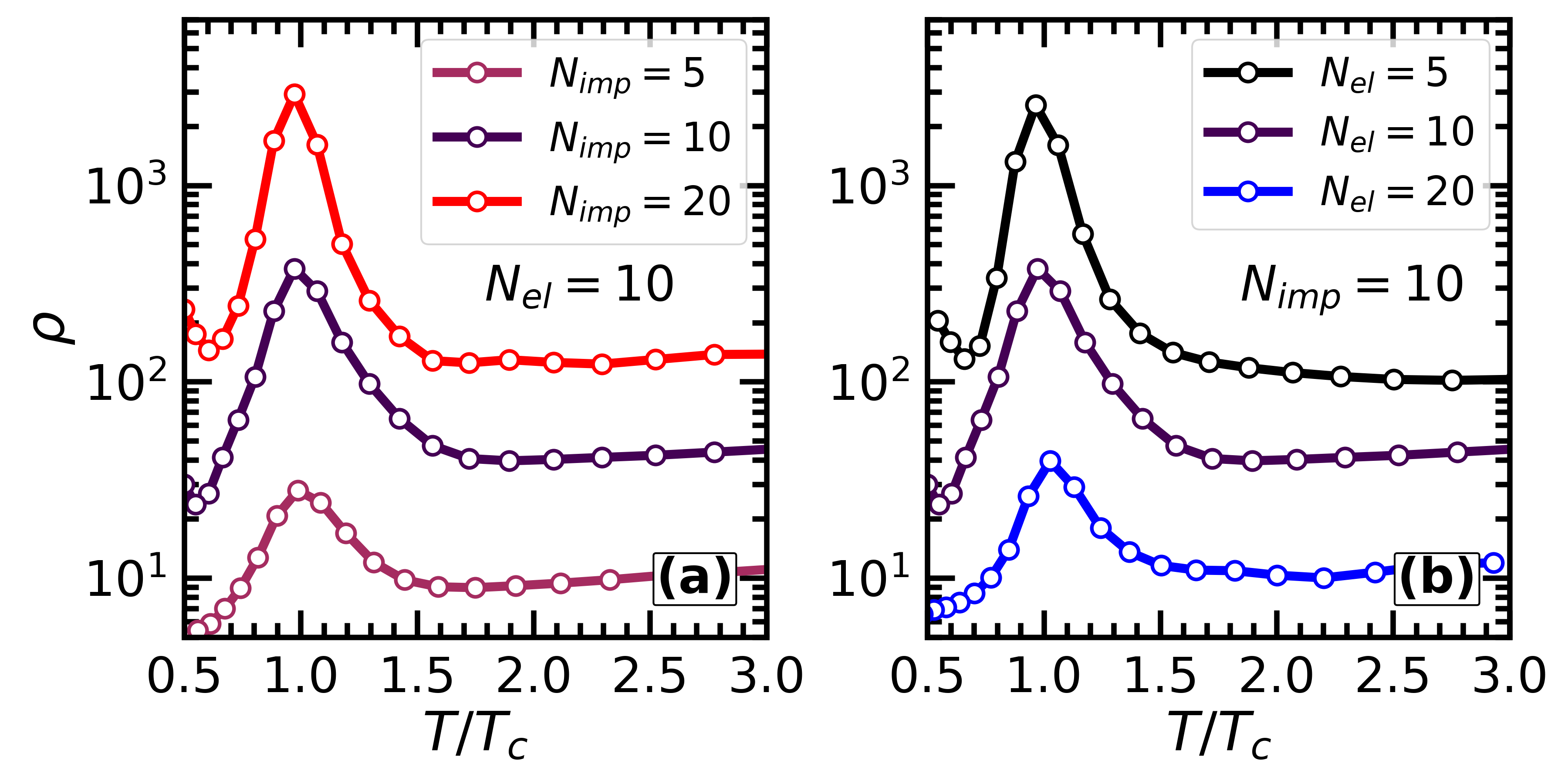} ~~ }
\caption{The resistivity computed for different combinations of
$N_{el}$ and $N_{imp}$ at $V=-2t$ to demonstrate that $\rho(T)$
depends essentially on the ratio $\eta = n_{imp}/n_{el}
= N_{imp}/N_{el}$ in the regime we are in.
}
\end{figure}

Having obtained $\epsilon_c$, we can now create a crude picture
of how $\epsilon_c - \mu$ changes with $T$. In the disorder
language if $\epsilon_c - \mu < 0$ the states at $\mu$
are extended, while for $\epsilon_c - \mu > 0$ the states
at $\mu$ are localised and the system is an insulator. Based
on the configurations available to us, we can do this systematically
for a dense grid of temperature, but we highlight the qualitative
behaviour based on only four $T$ points in Fig.10(b). 

At both $\eta = 0.5$ and $1.0$ $\epsilon_c$ rises from the
band bottom at some $T > 0.5T_c$, increases as $T 
\rightarrow T_c$ but does not cross $\mu$ and falls again 
as $T$ increases further. At $\eta = 2.0$ the mobility
edge is already above the band bottom at $0.5T_c$ and
at $T_c$ and $1.5T_c$ it is above the chemical potential.
In a 3D situation the disorder will have a perturbative 
effect on the low temperature resistivity, which will
rise $\propto \eta$, while the $T \sim T_c$ state will 
show a metal-insulator transition as $\eta$ drives the 
mobility edge across $\mu$! 

Since the non monotonic $T$ dependence of $\rho(T)$, and
it's increasing prominence with growing $\eta$, is the
most significant feature of these systems, one may want
some simple temperature dependent indicator 
that correlates with it.
The hint comes from the $A(\omega)$ plots in Fig.9,
where the $A$ at a given $\omega$ is a non monotonic
function of $T$. Rather than focus on a single $\omega$
we calculate $A_{av}$, the average spatial coverage 
of eigenstates within the energy window $\mu \pm T$ for
each $T$ and $\eta$. 
Panel (a) of Fig.11 shows $A_{av}$ for the 
zero-field case. For $\eta = 0.5$, unlike in clean 
systems, $A_{av}$ does not reach the full system area  
even at $T \ll T_c$; instead, it saturates at around 
$75\%$ of it because of the mild inhomogeneity in 
electrons density observed due to disorder.
It shows a minimum near $T_c$.
Similar trends are visible at the higher $\eta$ as well,
except the minimum values are progressively smaller.

Panel (b) shows the effect of the magnetic field on 
$A_{av}$ for $\eta = 1.0$. As seen in the previous section,
 increasing the field $h$ leads to a greater spatial spread of the 
 electrons, which is reflected in the behaviour of $A_{av}$. The
 suppression of the nonmonotonicity in $A_{av}$ correlates 
 with the weakening of the nonmonotonicity in $\rho(T)$ with 
 increasing $h$. Furthermore, the increase in $A_{av}$ at $T_c$
 with increasing $h$ indicates an enlargement of the polaron size, 
 consistent with the observation in panel (c) of Fig.7.

\subsection{`Scaling behaviour' with respect to $n_{imp}/n_{el}$}

We have discussed the effect of disorder
 using the parameter $\eta = n_{imp}/n_{el}$ while keeping the
 disorder strength fixed at $V = -2t$, based on the assumption
 that $\eta$ effectively captures the physics.
without having to bother about
$n_{imp}$ and $n_{el}$ separately. To
 validate this assumption, we analyze the resistivity $\rho(T)$
 by independently varying impurity number $N_{imp}$ at
 fixed electron number $N_{el}$ and vice versa, as shown,
respectively, in panels (a) and (b) of Fig.9.
 Panel (a) displays $\rho(T)$ for $N_{imp} = 5$, $10$, and
 $20$ with $N_{el} = 10$, while panel (b) shows $\rho(T)$
 for $N_{el} = 5$, $10$, and $20$ with $N_{imp} = 10$.
 A comparison of the panels shows a clear similarity of
 $\rho(T)$ when the values of $\eta$ are the same,
 irrespective of the values of $N_{el}$ and $N_{imp}$.

\subsection{Effect of dimensionality on $\rho$}

In 2D, due to the tendency to localise at arbitrary structural 
disorder, i.e, the absence of a mobility edge in the infinite 
volume limit, the low temperature resistivity we obtain on our 
finite lattices is quite large. Ideally, the $T \rightarrow 0$ 
resistivity would diverge as $L \rightarrow \infty$ at any 
finite $V$. That makes our low $T$ values unreliable, and the
large but finite $\rho$ we obtain at large $\eta$ leads to a
smaller value of $\rho(T_c)/\rho(0)$. This was one of the main
discrepancy between our 2D results and experimental data. 
To understand the effect of dimensionality on $\rho(T)$ we 
investigate the low temperature ($T \ll T_c$) and high 
temperature ($T \gg T_c$) resistivity for cubic lattice 
of size $10 \times 10 \times 10$. 

\begin{figure}[t]
\centerline{
\includegraphics[height=5.2cm,width=8.6cm]{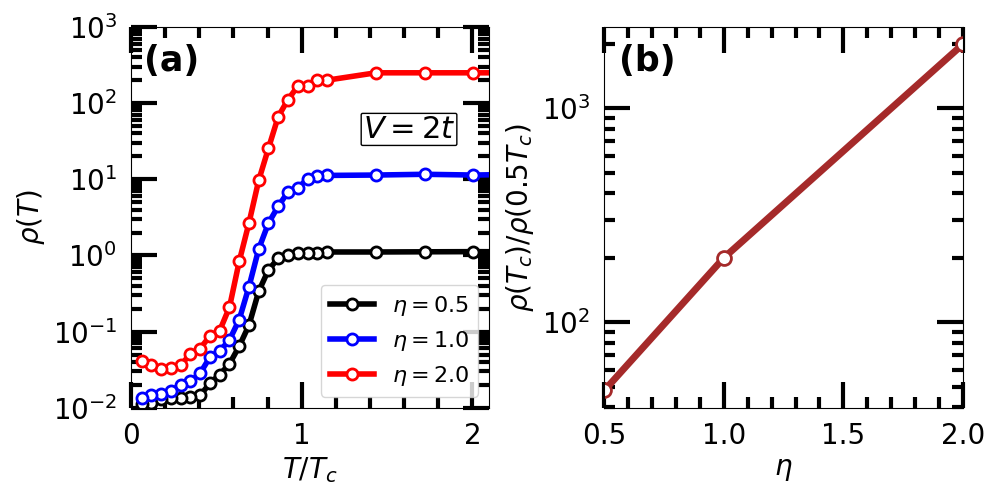}
~~
}
\caption{Resistivity in a 3D $10 \times 10 \times 10$ system where
the spin configurations are generated from a Heisenberg model
(with no electronic effects present) and the electron problem
diagonalised in these backgrounds with the impurity potential
present. The text explains the rationale, and our focus is
on the order of magnitude of $\rho(T_c)/\rho(0.5T_c)$ with
growing $\eta$. We use $V=-2t$.  (a)~$\rho(T)$ for three
values of $\eta$ showing a slow increase in low $T$
resistivity with $\eta$, an enormous increase near $T_c$,
and and an almost flat behaviour for $T \gtrsim T_c$.
(b)~Shows the ratio  $\rho(T_c)/\rho(0.5T_c)$, which 
we had calculated in 2D earlier. Compared to the 2D
magnitude of $\sim 10^2$ at $\eta = 2.0$, Fig.1(c), 
the 3D value is $\sim 2 \times 10^3$. This is without
any `polaronic effects', which we guess will enhance 
$\rho(T_c)$ further.
}
\end{figure}

Solving the polaron problem for this system size is impossible 
so
we look into a simpler problem. We generated spin configurations 
for the clean Heisenberg model via Monte Carlo and diagonalised 
the electronic Hamiltonian over the equilibrium spin configurations 
in the presence of the impurity potential. 
Such an approach retains the low $T$ disorder effects correctly
but misses the electron feedback on spin configurations that
can become prominent near $T_c$.
We think the $T \ll T_c$ resistivity would
be a fair representation of what one would see in the full
3D calculation. 
Also, for $T \gg T_c$, where spin entropy effects would 
dominate over polaron formation, one expects the resistivity
to be a reasonable mimic of the full problem. 

Fig.13(a) shows $\rho(T)$ for the usual values of $\eta$ and
we observe a slow growth of the low $T$ resistivity with
$\eta$ and a rapid growth of the $T \gtrsim T_c$ resistivity.
The change from low to high temperature value can span
three orders of magnitude. The relevant ratio $\rho(T_c)/\rho(0.5T_c)$
is shown in panel (b) and goes up to $ 2 \times 10^3$. This
is more than an order of magnitude larger than our
2D result in Fig.1(c). We believe that if polaronic effects
were included, the actual $\rho(T_c)$ in 3D would be larger.

\section{Conclusion}

We have studied the problem of ferromagnetic polaron formation in 
a structurally disordered background and compared our results to 
available data on EuO wherever possible. To our knowledge, the 
disordered polaron problem has never been addressed in a real 
space setting, handling both the structural disorder and the 
magnetic fluctuations exactly. We could accomplish this on a 
two dimensional $20 \times 20$ system by using a parallelised 
Langevin dynamics scheme where thermal noise brings in magnetic 
fluctuations while the impurity potential is handled through 
iterative exact diagonalisation. Probing the system as a function 
of increasing impurity density $(n_{imp})$, and fixed electron 
density $(n_{el})$, we find a non monotonic temperature dependence 
in the resistivity, with a peak near $T_c$ that grows by a 
factor of $O(10^4)$ 
from $n_{imp}=0$  to $n_{imp} = 2 n_{el}$. The growing anomaly near 
$T_c$ is also reflected in spatial localisation of electrons. The 
inhomogeneities become more pronounced with increasing $n_{imp}$ 
but show a non monotonic temperature dependence, with the sharpest
localisation near $T_c$. There is an associated pseudogap in the 
density of states and a shift of the optical conductivity to a non 
Drude form near $T_c$, both readily measurable features. We also 
established the magnetic field driven homogenisation and the 
associated huge magnetoresistance. Finally, we suggest and test 
out a scenario for electron localisation as a function of impurity 
concentration and temperature that should go beyond the specific 
2D numerics and parameter choice, and set a template for analysing
metal-insulator transitions in these complex materials.

\vspace{.2cm}

{\it Acknowledgment:} We acknowledge use of the High Performance
Clusters at HRI.

\section{Appendix}

\subsection{Computation of various indicators}

The primary output of the Langevin dynamics is a time series
for ${\bf S}_i(t)$, for each temperature and a given 
impurity configuration. 
Suppose each equilibrium spin configuration, that is used
for physical property calculation is indexed by a label
$\alpha$, i.e, the spin configurations are
$\{{\bf S}_i \}^{\alpha}$.
The single particle eigenvalues in this configuration 
would be $\epsilon_n^{\alpha}$ and the eigenstates would 
be $\psi^{\alpha}_n({\bf r})$.
The electron spin is not a quantum number in 
a generic ${\bf S}_i$ configuration so the index $n$
has $2N$ values where $N$ is the number of lattice sites.
We define the Fermi function
$f(\epsilon) = (e^{{\beta}(\epsilon - \mu)} + 1)^{-1}$.

We first write the expressions for various electronic
properties in a specific configuration $\alpha$. 
For most of the results, resistivity, optical conductivity, 
DOS, lDOS we have shown the results after configuration 
average, and we do so over
typically $100$ configurations. For, local density $n_i$ and
local spin correlation $C_i$ we show results for a 
single configurations without thermal averaging.

\begin{enumerate}
\item
Spatial density: 
\begin{equation}
n_{\alpha}({\bf r}) 
= \sum_n \vert \psi_n^{\alpha}({\bf r}) \vert^2 f(\epsilon_n^{\alpha})
\end{equation}
\item
The instantaneous local ferromagnetic correlation:
\begin{equation}
C_i=  \sum_{j}{\bf S}_i.{\bf S}_j,
\end{equation}
where $j$ runs over all the lattice points inside a circle centring $i$ 
and has a radius of a $3$ lattice spacing.
\item
Density of states.
\begin{equation}
D_{\alpha}(\omega) = \sum_n \delta(\omega - \epsilon_n^{\alpha})
\end{equation}
\item
Local density of states.
\begin{equation}
D_{\alpha}({\bf r}, \omega) = \sum_n 
\vert \psi_n^{\alpha}({\bf r}) \vert^2  \delta(\omega - \epsilon_n^{\alpha})
\end{equation}
\item
Spin resolved DOS: the LDOS above is spin summed. To know if there is a large
local magnetisation over a spatial neighbourhood it is useful to calculate
the spin resolved LDOS from the local Greens 
function $G^{\alpha}_{\sigma \sigma}({\bf r}, \omega)$.
\begin{eqnarray}
N^{\alpha}_{\sigma}({\bf r}, \omega) & = & - ({1 \over \pi}) Im 
[G^{\alpha}_{\sigma \sigma}({\bf r}, \omega)] 
\cr
\cr
G^{\alpha}_{\sigma \sigma}({\bf r}, \omega) &=&
\int_0^{\infty} 
dt e^{i \omega t} \langle \Psi^0_{\alpha} \vert 
[c_{{\bf r} \sigma}(t), c^{\dagger}_{{\bf r} \sigma}(0)]
\vert \Psi^0_{\alpha} \rangle
\end{eqnarray}
where $\vert \Psi^0_{\alpha} \rangle$ is the many body ground state
of the $N_{el}$ system (here we ignore Fermi factors).
\item
Optical conductivity.
\begin{eqnarray}
\sigma_{\alpha} (\omega) &=& A
\sum_{m,n} 
{ {f(\epsilon_m^{\alpha}) - f(\epsilon_n^{\alpha})}
\over {\epsilon_m^{\alpha} - \epsilon_n^{\alpha}} }
\vert {\hat j}_{mn}^{\alpha \alpha} \vert^2
\delta(\omega - ({\epsilon_m^{\alpha} - \epsilon_n^{\alpha}}))
\cr
{\hat j} &=& i t a_0 e \sum_{i, \sigma} (c^{\dagger}_{{i + x a_0},\sigma}
c_{i, \sigma} - h.c) 
\end{eqnarray}
where $A = {\pi e^2}/{\hbar N}$
\item
DC resistivity:
we obtained the `dc conductivity' $\sigma_{\alpha}$ as
\begin{equation}
\sigma_{\alpha} = {1 \over {\Delta \omega}} 
\int_0^{\Delta \omega} d \omega \sigma_{\alpha} (\omega)
\end{equation}
where $\Delta \omega$ is a small multiple of the average finite
size gap in the spectrum. In our calculations it is $0.05t$.
The d.c resistivity is the inverse of the thermally averaged
conductivity
$$
\rho = [ {1 \over {N_{\alpha}}} \sum_{\alpha} {\sigma_{\alpha}} ]^{-1}
$$
\item
Inverse participation ratio (IPR): this is useful to quantify the
inverse of the
`volume' associated with a single particle eigenstates. The standard
definition, for normalised states, is:
\begin{equation}
I^{\alpha}_n = \int d{\bf r} \vert \psi_n^{\alpha}({\bf r}) \vert^4
\end{equation}
D.c transport involves states over a window $\mu \pm k_BT$. Since the
character of states changes rapidly with energy in the polaronic
regime we calculate the typical area $A(T)$ associated with eigenstates
in the  $\mu \pm k_BT$ window as follows:
\begin{equation}
A_{\alpha}(T) = {1 \over N_S} \sum_n {1 \over {I^{\alpha}_n}}~,
~~~~~{\epsilon_n \in (\mu \pm k_BT)} 
\end{equation}
where $N_S$ is the number of states in the $\mu \pm k_BT$ window.
We then average $A_{\alpha}$ over configurations.
\end{enumerate}


\begin{thebibliography}{9}
\bibitem{footnote}
As system size $L \rightarrow \infty$ the conductivity at $T=0$ in our 
2D system will decrease at any finite disorder and tend to zero. This 
is a source of unreliability in our $T \rightarrow 0$ result vis-a-vis 
the 3D experiments. We comment on this in the Discussion section.
\end{thebibliography}
\end{document}